\documentclass[traditabstract]{aa} 
\usepackage{graphicx,lscape,natbib}
\usepackage{amssymb}
\usepackage{amsmath}
\usepackage{wasysym}
\usepackage{float}
\def\spb{\smallskip\par\noindent$\bullet\;$}
\def\lsim{\mathrel{\rlap{\lower4pt\hbox{\hskip1pt$\sim$}}
    \raise1pt\hbox{$<$}}}                
\def\gsim{\mathrel{\rlap{\lower4pt\hbox{\hskip1pt$\sim$}}
    \raise1pt\hbox{$>$}}}  
\usepackage{txfonts}
\font\aipsfont = cmsy9 scaled\magstep1

\newcommand\aips {{\aipsfont AIPS}}
\begin{document}
%
   \title{An excess of dusty starbursts related to the Spiderweb galaxy}


   \author{H. Dannerbauer\inst{1}, J.~D.~Kurk\inst{2}, C.~De Breuck\inst{3}, D.~Wylezalek\inst{3}, J.~S.~Santos\inst{4}, Y.~Koyama\inst{5,6}, N.~Seymour\inst{7}, M.~Tanaka\inst{5,8}, N.~Hatch\inst{9}, B.~Altieri\inst{10}, D.~Coia\inst{10}, A.~Galametz\inst{11},  T.~Kodama\inst{5}, G.~Miley\inst{12}, H.~R\"ottgering\inst{12}, M.~Sanchez-Portal\inst{10}, I.~Valtchanov\inst{10}, B.~Venemans\inst{13}, and B.~Ziegler\inst{1}}

   \institute{Universit\"at Wien, Institut f\"ur Astrophysik, 
              T\"urkenschanzstra\ss e 17, 1180 Wien, Austria\\
              \email{helmut.dannerbauer@univie.ac.at}
\and Max-Planck-Institut f\"ur extraterrestrische Physik, Giessenbachstra\ss e 1, 85748 Garching, Germany 
\and European Southern Observatory, Karl Schwarzschild Stra\ss e 2, 85748, Garching, Germany  
\and INAF-Osservatorio Astrofisico di Arcetri, Largo E. Fermi 5, 50125, Firenze, Italy
\and Optical and Infrared Astronomy Division, National Astronomical Observatory of Japan, Mitaka, Tokyo 181-8588, Japan
\and Institute of Space Astronomical Science, Japan Aerospace Exploration Agency, Sagamihara, Kanagawa 252-5210, Japan
\and  International Centre for Radio Astronomy Research, Curtin University, Perth, Australia
\and Kavli Institute for the Physics and Mathematics of the Universe, The University of Tokyo, 5-1-5 Kashiwanoha, Kashiwa-shi, Chiba 277-8583, Japan
\and School of Physics and Astronomy, University of Nottingham, University Park, Nottingham NG7 2RD, UK
\and Herschel Science Centre, European Space Astronomy Centre, ESA, 28691 Villanueva de la Ca\~{n}ada, Spain
\and INAF - Osservatorio di Roma, via Frascati 33, 00040, Monteporzio, Italy
\and Leiden Observatory, PO Box 9513, 2300 RA Leiden, the Netherlands
\and Max-Planck Institut f\"ur Astronomie, K\"onigstuhl 17, 69117 Heidelberg, Germany 
}

   \date{Received 7 March 2014; accepted XX 2014}

 
  \abstract{We present APEX LABOCA 870~$\mu$m observations of the field around
    the high-redshift radio galaxy MRC1138$-$262 at $z=2.16$. We detect 16
    submillimeter galaxies in this $\sim$140 arcmin$^{2}$ 
    bolometer map with flux densities in the range $3-11$~mJy. The
    raw number counts indicate a density of submillimeter galaxies (SMGs) that is up to four times that of blank field surveys. Based on an exquisite
    multiwavelength database, including VLA 1.4~GHz radio and infrared
    observations, we investigate whether these sources are members of the
    protocluster structure at $z\approx2.2$. Using {\it Herschel}
    PACS$+$SPIRE  and {\it Spitzer} MIPS photometry, we derive
    reliable far-infrared photometric redshifts for all 
    sources. Follow-up VLT ISAAC and SINFONI near-infrared spectra 
    confirm that four of these SMGs have redshifts of $z\approx2.2$. We also present evidence that another SMG in this field, detected earlier at 850 $\mu$m, has a counterpart that exhibits H$\alpha$ and CO(1-0) emission at z=2.15. Including the radio galaxy and two SMGs with far-IR photometric redshifts at z=2.2, we conclude that at least eight submm sources are part of the protocluster at $z=2.16$ associated with the radio galaxy MRC1138$-$262. We measure a star formation rate
    density SFRD~$\sim$1500~M$_{\odot}$~yr$^{-1}$~Mpc$^{-3}$, four magnitudes
    higher than the global SFRD of blank fields at this redshift. Strikingly,
    these eight sources are concentrated within a region of 2~Mpc (the
    typical size of clusters in the local universe) and are 
    distributed within the filaments traced
    by the H$\alpha$ emitters at $z\approx2.2$. This concentration of
    massive, dusty starbursts is not centered on the submillimeter-bright radio galaxy which could support the infalling of these sources into the cluster center. Approximately half (6/11) of the SMGs
that are covered by the H$\alpha$ imaging data are associated with
    H$\alpha$ emitters, demonstrating the potential of tracing SMG
    counterparts with this population. To summarize, our results demonstrate that submillimeter 
    observations may enable us to study (proto)clusters of massive, dusty
    starbursts.}


\authorrunning{H.~Dannerbauer et al.}
\titlerunning{APEX LABOCA observations of the field around MRC1138$-$262}

\maketitle
%

\section{Introduction}
The questions of when and how present-day galaxy clusters formed at high
redshift have driven extensive searches for
protoclusters of galaxies in the distant Universe in the past two decades
\citep[e.g.,][]{lef96,ste98,pen00,kur00,kur04a,kur04b,bes03,mat05,dad09a,gal10,gal12,hat11a,hat11b,may12,wal12,wyl13}. Powerful
high-redshift radio galaxies \citep[HzRGs; see the review by
  ][]{mil08} are considered to be the most promising
signposts of the most massive clusters in formation.  Surveys
of Ly$\alpha$ emitters (LAEs), H$\alpha$ emitters (HAEs), Lyman break
galaxies (LBGs), and extremely red objects (EROs) in several fields
containing radio galaxies, up to redshifts of 5.2, produced evidence of galaxy overdensities in almost
all cases \citep[e.g.,][]{kur00,mil06,ove06,pen00,ven02,ven04,ven05,ven07}, even
out to 10 Mpc \citep{int06}. 
These surveys convincingly
demonstrate that HzRGs are good signposts of 
overdensities of galaxies at high redshift, at least in optical and near-infrared
bands.

In the past decade (sub)millimeter surveys have revolutionized our
understanding of the formation and evolution of galaxies by revealing
a population of high-redshift, dust-obscured galaxies
that are forming stars at a tremendous rate.  Submillimeter galaxies
\citep[SMGs; see the review by][]{bla02}, first discovered by
\citet{sma97}, have intense star formation, with rates of a few
hundred to several thousand solar masses per year, but due to strong dust obscuration inconspicuous at optical/NIR wavelengths \citep[e.g.,][]{dan02,dan04}. These dusty
starbursts are massive \citep[a few times $10^{11}$~M$_{\odot}$, see e.g.,][]{gen03,gre05}, and are probably the
precursors of present-day ellipticals
\citep[e.g.,][]{lut01,ivi13}. Furthermore, SMGs are not uniformly
distributed \citep{hic12} and are excellent tracers of mass density peaks \citep{oui04}
and thus of so-called protoclusters --- the precursors of structures
seen in the local universe such as the Coma cluster. These early
(proto)clusters place significant constraints on models of galaxy
assembly at those redshifts \citep{ster10}, and offer us a unique opportunity to
explore episodes of bursting star formation in a critical epoch of
galaxy formation. 

Up to now, large scale structures like overdensities
of galaxies have only been found through optical/near-infrared
observations. However, we note that these optical and NIR techniques
mainly trace (rather low-mass) galaxies with unobscured star
formation, making up only 50\% of the cosmic star formation activity
\citep{dol06}. As outlined above, overdensities of unobscured star
forming galaxies have been detected around a significant sample of
HzRGs, but the detection of obscured star forming galaxies in these
fields is lagging behind. Several studies report an excess of SMGs near HzRGs and QSOs
\citep[e.g.,][]{ste03,deb04,gre07,pri08,ste10,car11,rig14}. However,
the recent analysis of {\it Herschel} observations of the field of
4C$+$41.17 at $z=3.8$ by \citet{wyl13} illustrates the importance of
determining the redshifts of the SMGs. \citet{wyl13} show that most of the {\it Herschel} sources
are foreground to the radio galaxy, casting doubts on the earlier
claim from \citet{ivi00} of an overdensity related to the radio galaxy based on SCUBA
observations.

One of the best studied large scale structures so far is the
protocluster associated with the HzRG MRC1138$-$262 at
$z=2.16$, the so-called Spiderweb galaxy \citep{mil06}. Ly$\alpha$ and H$\alpha$ imaging/spectroscopy of this field reveal an
excess of LAEs compared to blank fields
\citep{kur00,pen00,kur04a,kur04b,hat11b}. Two attempts to search for
submillimeter overdensities on this field are known. Using SCUBA,
\citet{ste03} report the (tentative) excess of SMGs, and spatial
extension of the submillimeter emission of the HzRG MRC1138$-$262. However, we
note that the field of view of SCUBA only has a diameter of
2$^{\prime}$ ($\sim$1~Mpc at $z=2.16$), and thus the reported SMG excess is based
on very small numbers. \citet{rig14} present {\it Herschel} SPIRE
observations of a larger field ($\sim$400~arcmin$^{2}$), centered on
the HzRG. They report an excess of SPIRE~500~$\mu$m sources but found
no filamentary structure in the far infrared as seen in the rest-frame
optical \citep{kur04a,koy13a}. However, in both cases no counterpart
identification was attempted for the individual sources. In addition,
\citet{val13} report the serendipitous discovery of an overdensity of
SPIRE 250~$\mu$m sources 7$^{\prime}$ south of the
protocluster. Based on the modfied blackbody derived redshift
distribution, incorporating both the color information and the SED
shape, they conclude that the majority of the 250~$\mu$m sources in
the overdensity are likely to be at a similar redshift. With the
available scarce multiwavelength data they cannot exclude the
attractive possibility that the overdensity is within the same
structure as the Spiderweb at $z\approx2.2$.

In this paper, we present our search for SMGs in the field of
MRC1138$-$262 using APEX LABOCA 870~$\mu$m observations. We discover
16 LABOCA sources, which is a significant excess of SMGs compared to
blank field surveys. We identify the counterparts of the SMGs using
the existing exquisite multiwavelength data on this field
\citep[][]{pen00,kur04a,kur04b,sey12,koy13a}. The main aim of this work is
to verify how many of the 16 SMGs are part of the well-known
protocluster structure at $z\approx2.2$. We mainly focus on {\it Spitzer}
MIPS, {\it Herschel} PACS$+$SPIRE, and VLA data, complemented by narrow-band images of H$\alpha$ emitters (HAEs) at $z\approx2.2$. We show that
H$\alpha$ emitters can readily be used to identify the counterparts of
SMGs since several SMGs are bright in H$\alpha$.  

The structure of this
paper is as follows. Sections 2 and 3 describe the observations of the
field around the HzRG MRC1138$-$262 and the associated LABOCA
sources. In Section 4 we present the method for deriving far-infrared
photometric redshifts, luminosities and star formation rates for the
LABOCA sources. In Section 5 we discuss the sources individually and
in Section 6 the properties of the SMG overdensity. We adopt the
cosmological parameters $\Omega_{matter}=0.27$,
$\Omega_{\Lambda}=0.73$, and $H_{0}=71$~km~s$^{-1}$~Mpc$^{-1}$
\citep{spe03,spe07}.


\section{Observations and data reduction}
\subsection{LABOCA imaging}
We mapped a field of $\sim$140~arcmin$^{2}$ around the HzRG
 MRC1138$-$262 with the bolometer camera LABOCA \citep{sir09}
installed on the APEX telescope through ESO (ID: 084.A-1016(A), PI:
J.~D. Kurk) and Max-Planck-Gesellschaft (MPG, ID: 083.F-0022, PI: J.~D. Kurk) time. The LABOCA
instrument contains 295 bolometer elements and operates at an
\begin{figure}[!b]
 \centering
 \includegraphics[width=8.cm,angle=0]{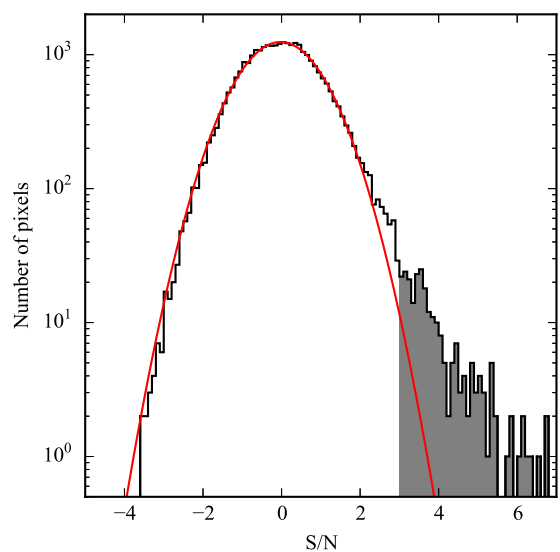}
 \caption{Histogram of the pixel
   signal-to-noise values of our LABOCA map. The red solid line shows
   a gaussian fit. The significant excess of pixels with a S/N~$\ge 3\sigma$ is
   shown in grey. Almost all of these pixels belong to selected sources and none to pixels at the edge of the map.}
 \label{fig:snrhistogram}
\end{figure}
effective frequency of 345~GHz corresponding to 870~$\mu$m. The LABOCA
array covers a field of view of 11.4$^{\prime}$ with a FWHM of
19$^{\prime\prime}$ at 870~$\mu$m. The observations were taken between
August and December 2009 in service mode, under excellent atmospheric
conditions with typical zenith opacities between 0.2 and 0.3 at
870~$\mu$m. The total on sky integration time was 16.6~hours.

We used the raster spiral scanning mode which combines the spiral scanning pattern with raster mapping. This mode has the advantage
of producing a fully sampled map of the total field-of-view of LABOCA
in a dense sampling pattern. The calibration observations were performed on a regular
basis and included
pointing, focus and flux calibration, see \citet{sir09} for more details.  Each scan was carefully
inspected for the presence of possible outliers, anomalies, and the 
influence of instabilities in the atmosphere. The data were reduced
using {\tt miniCrush} \citep{kov08}, a commonly used software for the
reduction of (sub)millimeter bolometric data. We used the option {\it
  \lq-deep'} 
 that is optimized for the reduction of deep field data containing
faint, point-like sources. The end product of the {\tt miniCrush} reduction is a multi-frame FITS image
containing a signal map, a noise map, a signal-to-noise map and an
exposure time map.

\begin{figure*}[!]
 \centering
 \includegraphics[width=18cm,angle=0]{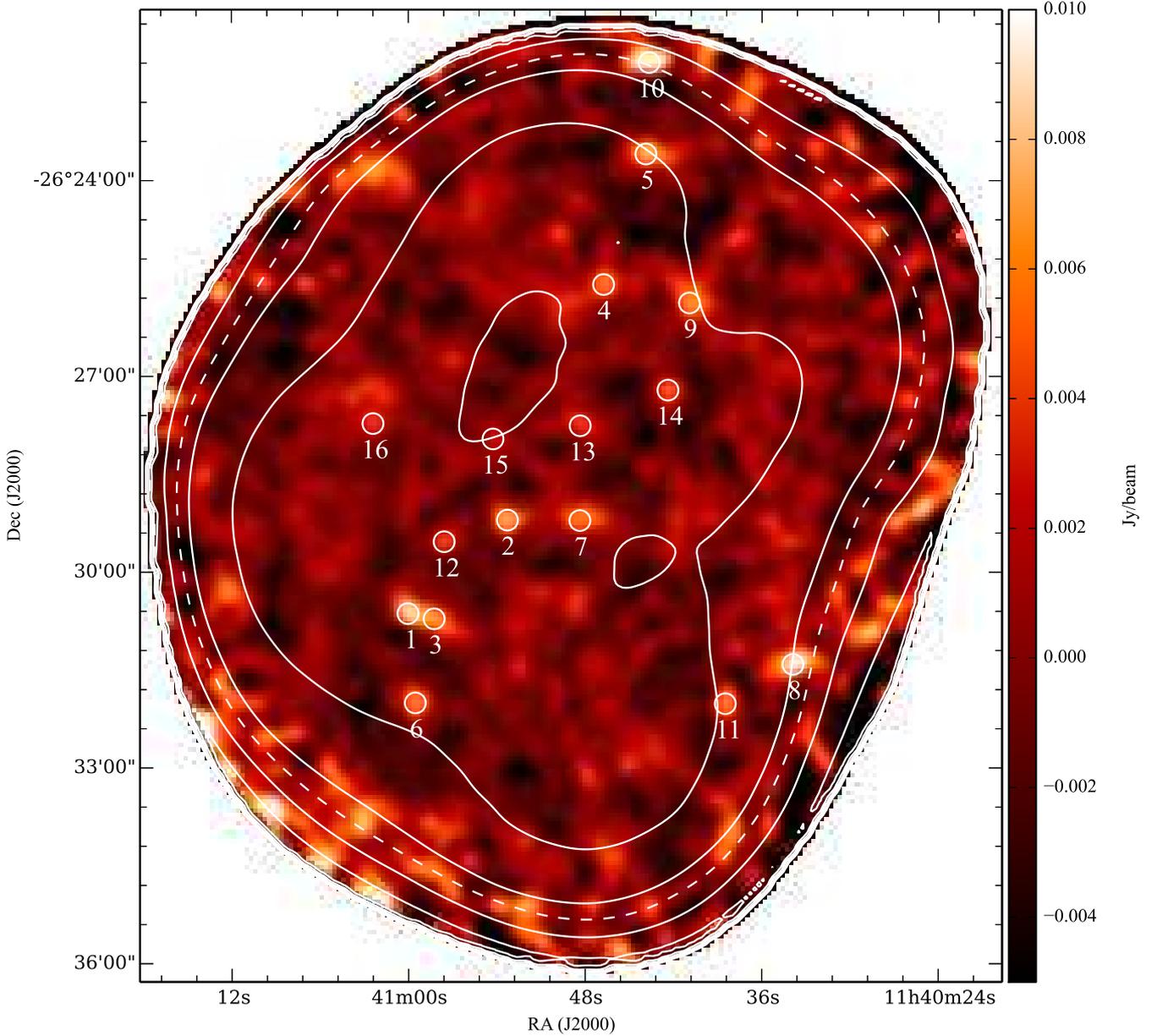}
 \caption{LABOCA signal map of the field around the high-z radio
   galaxy MRC1138$-$262 (\# 7). Encircled, we show the location of 16 SMGs
   extracted from our LABOCA map. Contours indicate the noise at 1.3,
   1.9, 2.6, 3.0, 3.7, 5.2 and 7.4~mJy/beam. The dashed contour encompasses the region where the noise is $\sigma<$~3.0~mJy/beam, including all selected 16 LABOCA sources. The source density is
   up to 4$\times$ higher than in the ECDFS. North is at the top and east is
   to the left.}
 \label{fig:labocasignal_noise}
\end{figure*}

\begin{table*}
\begin{center}
\caption{870~$\mu$m LABOCA Source Catalog of the Field of MRC1138$-$262}
\label{tab:laboca}
\begin{tabular}{cccccc}
Source&Alias&R.A.&Decl.&$S_{870 \mu m}$&S/N\\
(IAU)&&(J2000.0)&(J2000.0)&(mJy)\\
(1)&(2)&(3)&(4)&(5)&(6)\\
\hline\hline
SMM~114100.0$-$263039 & DKB01 & 11:41:00.04 & $-$26:30:39.2 & 9.8$\pm$1.5 &6.7\\
SMM~114053.3$-$262913 & DKB02 & 11:40:53.28 & $-$26:29:14.0 & 8.1$\pm$1.5 &5.4\\
SMM~114058.3$-$263044 & DKB03 & 11:40:58.26 & $-$26:30:44.0 & 7.3$\pm$1.5 &4.9\\ 
SMM~114046.8$-$262539 & DKB04 & 11:40:46.75 & $-$26:25:39.2 & 6.8$\pm$1.4 &4.7\\
SMM~114043.9$-$262340 & DKB05 & 11:40:43.88 & $-$26:23:40.2 & 8.2$\pm$1.8 &4.5\\
SMM~114059.5$-$263200 & DKB06 & 11:40:59.54 & $-$26:32:00.7 & 6.8$\pm$1.7 &3.9\\
SMM~114048.4$-$262914 & DKB07 & 11:40:48.36 & $-$26:29:14.4 & 6.7$\pm$1.7 &3.9\\
SMM~114033.9$-$263125 & DKB08 & 11:40:33.88 & $-$26:31:25.6 & 10.6$\pm$2.7 &3.9\\
SMM~114040.9$-$262555 & DKB09 & 11:40:40.92 & $-$26:25:56.0 & 7.1$\pm$1.9 &3.8\\
SMM~114043.7$-$262216 & DKB10 & 11:40:43.66 & $-$26:22:16.8 & 11.0$\pm$3.0 &3.7\\
SMM~114038.5$-$263201 & DKB11 & 11:40:38.48 & $-$26:32:01.4 & 7.0$\pm$1.9 &3.6\\
SMM~114057.6$-$262933 & DKB12 & 11:40:57.58 & $-$26:29:33.7 & 5.0$\pm$1.4 &3.6\\

\hline
cross-identified tentative detections\\
SMM~114048.3$-$262748 & DKB13 & 11:40:48.34 & $-$26:27:48.0 & 4.4$\pm$1.5 &3.0\\ 
SMM~114042.4$-$262715 & DKB14 & 11:40:42.38 & $-$26:27:15.5 & 5.3$\pm$1.8 &3.0\\
SMM~114054.3$-$262800 & DKB15 & 11:40:54.26 & $-$26:28:00.0 & 3.2$\pm$1.3 &2.4\\
SMM~114102.7$-$262746 & DKB16 & 11:41:02.41 & $-$26:27:46.0 & 4.2$\pm$1.4 &2.9\\
\hline
\end{tabular}
\tablefoot{--- Col.~(1): LABOCA source. Col.~(2): Short name of LABOCA source. Col.~(3): J2000.0 right ascension of LABOCA source. Units of right ascension are hours, minutes, and seconds. Col.~(4): J2000.0 declination of LABOCA source. Units of declination are degrees, arcminutes, and arcseconds. Col.~(5): LABOCA flux. Col.~(6): Signal-to-Noise of LABOCA detection.}
\end{center}
\end{table*}

In the central part of the LABOCA map ($\sim$56~arcmin$^{2}$), we
achieve an rms noise level of 1.3--1.9~mJy. In Fig.~\ref{fig:snrhistogram}, we
show the pixel signal-to-noise distribution of our LABOCA map. The main distribution of pixel values are well fit by a Gaussian centered at zero.  However,
there is significant excess of positive valued pixels with  a signal-to-noise ratio (S/N) $\ge 3\sigma$. This skewed distribution indicates that the pixels with excess are associated with real submillimeter sources.  We have checked this by identifying
all pixels with a S/N $\ge 3\sigma$ on the map and confirmed that almost all of
these belong to sources identified in Section \ref{sec:notes} and none lie near the edge
of the map.

We have searched our LABOCA signal-to-noise map within the region where the noise is $\sigma<3.0$~mJy/beam (Fig.~\ref{fig:labocasignal_noise}) for S/N peaks
down to 3.5$\sigma$. Furthermore, we cross-identified LABOCA S/N peaks below 3.5$\sigma$ which are detected at similar submm wavelengths, in our case by {\it Herschel} (see below this and subsection \ref{sec:pan} for more details) as potential LABOCA sources. The detected sources had to have at least the size of the LABOCA beam. In Table~\ref{tab:laboca} we list all 16 sources in order of signal-to-noise and from now on we use their alias (DKB01--DKB16). Twelve LABOCA sources are classified as secure (S/N$\geq3.5$), the remaining four sources are classified as cross-identified tentative. We give the position of the pixel with the highest S/N and list the peak fluxes --- a standard technique in radio astronomy -- obtained from the signal map. We used two approaches to verify the reliability of our selected sources. The first approach relies on checking observations of the same field at similar wavelengths. We use our {\it Herschel} PACS$+$SPIRE dataset, see forthcoming section~\ref{sec:pan} for more details. With these deep and wide 'auxiliary' {\it Herschel} data, we can very well discard spurious LABOCA sources. Only one out of 16 sources (DKB09) is not detected at any of the {\it Herschel} bands. Especially, all of the four 'tentative' LABOCA sources have significantly detected {\it Herschel} counterparts, see also Table~\ref{tab:id}. The second approach is based on the so-called jackknife technique: We split the ESO and MPG data into two groups of similar integration time. All 16 sources were detected in both datasets.  Finally, we investigated the reliability of our source extraction approach. For this sanity check, we used the source extraction tool {\tt detect}, part of the software package {\tt Crush} \citep{kov08}. Beside the $2.4\sigma$ source, all sources could be „recovered“ by this extraction algorithm down to 3.0$\sigma$, giving us faith in our approach. For sources with $\gsim3.5\sigma$, the false detection rate is estimated on 0.2 sources among the 12 secure sources, justifying that we call this sample „secure“. To guarantee a proper comparison with the only known LABOCA deep field on the ECDFS \citep[LESS][]{wei09}, we used only our 3.7$\sigma$ sources for surface density calculations.


%

\subsection{VLA imaging}
The MRC~1138-262 field was observed with the Karl G. Jansky Very Large Array
\citep[VLA;][]{nap83} on UT 2002 April 1-12 for a total of
12~hours in A configuration at 20~cm (ID: AD0463, PI: C. De Breuck). We observed in a pseudo-continuum,
spectral line mode with $7 \times 3.125$~MHz channels. The point source 1351$-$148 was monitored every 40~min to obtain amplitude, phase
and bandpass calibration, and an observation of 3C~286 was used to obtain 
the absolute flux calibration.

Standard spectral-line calibration and editing of the
data was performed using the NRAO \aips\ package and standard wide field
imaging techniques \citep{tay99}. The final 7\farcm5$\times$7\farcm5
image has an rms noise level of 19~$\mu$Jy~beam$^{-1}$, except in an
area close to the central radio galaxy which is limited by the
ability to clean the bright radio source. The dynamic range achieved
is $\sim$10$^4$. The FWHM resolution of the restoring beam is
$2\farcs7 \times 1\farcs3$ at a position angle PA=-10\degr.

\subsection{Panchromatic observations}
\label{sec:pan}
To analyse the 16 LABOCA sources we used several additional data sets
(see also Fig.~\ref{fig:member}):

\spb {\it H$\alpha$ spectroscopy:} In February 2012 we conducted
VLT ISAAC long-slit near-infrared spectroscopic observations (ID:
088.A-0754(A), PI: J.~D. Kurk) of the redshifted $H\alpha$ line in
visitor mode in order to confirm the redshifts of several LABOCA sources that were likely to be 
protocluster members. A detailed
discussion of these observations will be
presented in Kurk et al. (in prep.). In the current paper we 
only use the redshifts from these NIR spectra for our analysis and discussion. Furthermore, very
recently we obtained VLT SINFONI IFU spectrosopy (ID: 090.B-0712(A), PI: J.~D. Kurk)
data of four likely merging galaxies at $z\approx2.2$ that are in the LABOCA
FWHM of one of the SMGs.

\spb{\it H$\alpha$ imaging:} A total area of $\sim$50~arcmin$^2$
was imaged with the MOIRCS camera on the SUBARU telescope using a narrow-band filter covering H$\alpha$ emitted at the redshift of the
radio galaxy  \citep[for
  a detailed description see][]{koy13a}. The narrow-band filter
NB2071 ($\lambda=2.068~\mu$m, $\Delta\lambda=0.027~\mu$m) covered the
redshift range $z=2.13-2.17$. These data encompassed the smaller
($\sim$12~arcmin$^{2}$) but deeper H$\alpha$ data taken with the
VLT ISAAC by \citet{kur04a,kur04b}. 

\spb{\it Ly$\alpha$ imaging:} A subsection of the LABOCA field
($\sim$49~arcmin$^{2}$) was imaged in Ly$\alpha$ redshifted to
$z=2.16$. Details of these observations can be found in \citet{kur00,kur04a}.

\spb{\it Herschel data:} This field was observed in the far-IR with the
instruments PACS and SPIRE onboard of the {\it Herschel} Space Observatory  \citep{pil10,pog10,gri10}
as part
of the project scientist guaranteed time (PI: B. Altieri). These observations are presented in detail in 
\citet{sey12} and \citet{val13}.   The PACS images achieve
$3\sigma$ sensitivities of $\sim$4.5~mJy and $\sim$9.0~mJy at
100~$\mu$m and 160~$\mu$m, respectively. The SPIRE images
achieve $3\sigma$ sensitivities of $\sim$7.5~mJy, $\sim$8.0~mJy and
$\sim$9.0~m Jy at 250~$\mu$m, 350~$\mu$m and 500~$\mu$m,
respectively. The size of the PACS and SPIRE maps are
$\sim$120~arcmin$^{2}$ and $\sim$900~arcmin$^{2}$ respectively
\citep{val13}. The entire LABOCA map is covered by the SPIRE data and almost completely covered by PACS. There is also a wider SPIRE map ($\sim$30 arcmin radius) of similar depth that is presented in \citet{rig14}, however the extended regions are not required for the present analysis.

\spb{\it Spitzer MIPS 24~$\mu$m imaging:} We use archival {\it Spitzer}
5$^{\prime}\times$5$^{\prime}$ MIPS 24~$\mu$m images which are centered on
the HzRG and cover about 20\% of the LABOCA image. These data are used to derive far-IR photometric redshifts and establish the SEDs of the LABOCA
counterparts.

\spb{\it Optical/near-infrared photometric redshifts:} \citet{tan10} derived photometric redshifts based on $UgRIzJHK_{s}$ and three IRAC channel photometry of the field
covered by {\it Spitzer}. 

\section{LABOCA source counterparts}
We searched for LABOCA counterparts  within $9\farcs5$ of the
LABOCA detection in the MIPS~24~$\mu$m, PACS~100/160~$\mu$m and 
VLA~1.4~GHz images (see Table~\ref{tab:id}). These wavelengths (in particular 1.4\,GHz) are well suited for finding SMG counterparts and obtaining more precise positions than measured by the bolometer data \citep[e.g.,][]{dan04,dan10,pop06}. The search circle is consistent with the FWHM of the
LABOCA beam and guarantees that no reliable associations are
missed. For each candidate counterpart within the
search radius we calculate the corrected Poissonian probability {\it
  p} that the SMG association is a chance coincidence. This approach
 corrects the simple Poissonian probability of a
detected association for the possibility of associations of different
nature but similar probability \citep[][]{dow86} and is widely applied and accepted in the community \citep[e.g.,][]{ivi02,dan04,dan10,big11,sma14}.  It basically depends on the search radius, the distance of the potential counterpart to the LABOCA source and the source surface density down to the flux level of the potential LABOCA counterpart. More recent work that uses this method for SMGs can be found e.g. in \citet{big11}. 

Several previous attempts to locate secure
counterparts of SMGs have been done using optical and
near-infrared broad-band images \citep[e.g.,][]{web03b}, the most successful one is to use {\it Spitzer} IRAC data
\citep[e.g.,][]{pop06,hai09,big11}. However, finding
counterparts by applying p-statistics on optical and near-infrared
images has not been very successful \citep[e.g.,][]{web03b}. This is primarily because of the high surface density of (faint) optical and near-infrared
sources which are not associated to the far-IR and
radio sources. Applying the p-statistic method is most
promising using data with low surface densities of sources, such as radio or far-IR
images. In addition to this we find counterparts to several LABOCA sources in the H$\alpha$
imaging data. This motivates us to test a new approach by applying the
p-statistic method to the H$\alpha$ emitters in the field of MRC1138$-$262
\citep{koy13a}; cf. with \citet{sma14} who associated [OII] emitters successfully in the field and at the redshift of the cluster Cl0218.3$-$0510 with SMGs selected in this region.

Due to the large beam
size of the SPIRE data (FWHM$=18^{\prime\prime}-36^{\prime\prime}$) there is a large uncertainty in the measured position and source blending is a big problem. We therefore do not apply the p-statistic method to the SPIRE data. The derived probabilities of PACS, MIPS
and H$\alpha$ emitter (HAE) associations are based on raw number counts in
the LABOCA field. At all wavelengths we search for counterparts of
SMGs down to a signal-to-noise of $3\sigma$. Bright radio emission from MRC1138 causes spurious sources in the VLA map that can not be "cleaned". We therefore decide to assess the reliability of VLA counterparts
using published number counts from e.g. \citet{fom06}. Similar to e.g. \citet{dan10}, we define the following quality criteria for
assessing the robustness of identified candidate counterparts: we classify SMG associations with $p \leq
0.05$ as secure, and those with $0.05 < p \leq 0.10$ as possible or tentative
counterparts. Below we briefly discuss the results of the associations at
different wavelengths.

\subsection{VLA 1.4.~GHz counterparts}
Due to the limited dynamic range caused by the strong emission of the
radio galaxy itself, the VLA map is shallow in comparison to
other deep VLA integrations of submillimeter fields \citep[e.g.,][]{mor10}. Eight out of 16 LABOCA sources, including the radio galaxy, have a VLA
counterpart down to $3\sigma$. The shallow depth of the VLA map may explain the rather low identification rate of 50\%, however it is consistent with
previous identification rates which range between 40 to 60\%
\citep[e.g.,][]{ivi02,dan04}. There are no cases of
multiple VLA counterparts to a single LABOCA source in the field of MRC1138$-$262 and all VLA associations are classified as secure counterparts with fluxes between 60 and
620~$\mu$Jy. However, we note that mm-interferometric observations 
have recently shown that not all VLA sources within the (sub)mm
bolometer beam produce (sub)mm emission
\citep[e.g.,][]{you07,bar12,kar13,hod13}, so we must be cautious in the interpretation of the eight VLA counterparts.

\subsection{PACS~100 and 160~$\mu$m counterparts}
We uncover PACS counterparts for nine SMGs,
corresponding to a PACS identification rate of 56\% of our whole SMG sample
(see Table~\ref{tab:id} for details). We detect 9 (6) SMGs at 160 (100) $\mu$m. All
PACS~100~$\mu$m LABOCA associations are also detected at 160\,$\mu$m. Although the PACS data are
shallower than that of the PEP data of GOODS-N \citep{lut11}, we find a higher identification rate than that reported by \citet{dan10} for the
GOODS-N field. This could indicate that a significant fraction
of LABOCA sources are at redshift $z=2.16$, a redshift 
still accessible by {\it Herschel} PACS \citep[see e.g. Fig.~3
  in][]{dan10}. The 160~$\mu$m measurements lie
close to the far-IR peak so it is unsurprising that the number of PACS counterparts at 160~$\mu$m
is higher than at 100~$\mu$m \citep[see also][]{dan10}.  We note that based on the corrected Poissonian
probability $p$, each PACS detection within the bolometer beam (our
search circle) is classified as an associated SMG counterpart, being
consistent with the results reported by \citet{dan10}. PACS
fluxes of these dust-obscured sources range between 5.2~mJy to
530.9~mJy at 100~$\mu$m and 15.7~mJy to 652.0~mJy at 160~$\mu$m. Due to
the shallowness of the VLA data, two PACS counterparts are not
detected in the radio regime.

\subsection{MIPS 24~$\mu$m counterparts}
Due to the high surface density of MIPS 24~$\mu$m sources compared to
VLA or {\it Herschel} sources the p-statistic is not as useful as in
the radio or far-infrared regime. However, for completeness we
performed the p-statistic for MIPS sources as well. In total, seven
LABOCA sources are covered by the {\it Spitzer} MIPS 24~$\mu$m
map. Except in one case (DKB13), VLA/HAE counterparts are
associated with MIPS 24~$\mu$m sources. However, only three of them are classified as secure. In two cases the MIPS 24~$\mu$m detections would not have been classified as statistical possible associations demonstrating the very limited use of p-statistic applied on this source population.


\subsection{H$\alpha$ emitting counterparts}
\label{sec:hae}
Eleven~LABOCA sources are covered by the map of H$\alpha$ emitters at
$z=2.16$, of which seven SMGs (DKB01, DKB03, DKB07, DKB08, DKB12, DKB16,
DKB15) have H$\alpha$ emitters within their LABOCA beams. In three
cases, DKB01, DKB07 and DKB12, we find two, three and four HAEs within
the search radius, respectively.  Thirteen HAEs from \citet{koy13a} are
LABOCA counterparts of which 10 are classified as robust
counterparts and only three are tentative. Therefore, this seems
to be a very promising approach in order to find SMG counterparts (with
subarcsecond position accuracy) for a (proto)cluster with known redshift.

13 out of 83 H$\alpha$
emitters in the field of MRC1138 are correlated with
SMGs. \citet{koy13a} report 15 MIPS~24~$\mu$m associations (classified as dusty H$\alpha$ emitters) out of 60
HAEs in the MIPS FOV of 5$^{\prime}$$\times$5$^{\prime}$.  Interestingly, five out of 13 HAEs (covered
by MIPS) have MIPS~24~$\mu$m counterparts. The higher rate of MIPS associations is consistent with SMGs being dusty. Except for the
HzRG, only two SMG counterparts identified as a HAE coincides with a
robust VLA counterpart. Recent rest-frame H$\alpha$ spectroscopy
(Kurk et al., in prep.) confirms the redshifts of seven~HAEs which are
associated with LABOCA sources.

The {\it Herschel} PACS detections at 100 and/or 160~$\mu$m of all secure HAEs
associated with LABOCA sources except one, reinforces our hypothesis that these sources
are the true SMG counterparts. In the case of DKB12 all four
HAEs are within/at the edge of the PACS~160~$\mu$m detection (no
detection at 100~$\mu$m). None of the remaining HAEs classified as
tentative SMG counterparts and without redshift confirmation are
detected by PACS. We
primarily use the H$\alpha$ imaging data from \citet{koy13a} for our analysis, however,
to be complete we check if there are more HAEs in the deeper data of
\citet{kur04a}.  

\section{Far-infrared photometric redshifts, luminosities and star formation rates}
\label{sec:firprop}
We derive far-infrared photometric redshifts of the LABOCA sources to test the hypothesis of how many of the 16 LABOCA sources are part of the protocluster structure at $z=2.2$. Since the launch of {\it Herschel} far-IR photometric redshift determination has been established as a reliable diagnostic tool in order to investigate SMGs \citep[e.g.,][]{amb10,ros12,pea13,swi14}. Our far-infrared photometric redshifts are calculated using the code {\tt
  hyperz} \citep*{bol00} which minimizes the reduced $\chi^{2}$ to
find the best photometric redshift solution.

We use both synthetic and
empirical AGN and starburst templates from the SWIRE template library
\citep{pol07} complemented with self-constructed SED templates. The
latter are obtained by spline interpolation of the mid- and far-infrared
emission from LABOCA sources with confirmed spectroscopic
redshifts. The far-IR emission of 
submillimeter sources, particularly those at high redshift, will be a superposition of the emission from stellar-heated dust and AGN activity. In most cases SED template libraries are
derived from low-redshift sources and therefore often fail in
fitting the far-IR dust-bump for high redshift sources, whose shape
can vary greatly due to differing contributions from starburst and
AGN \citep[e.g.,][]{lag05,pol07,ski11}. 

We choose sources DKB07, 12c, 13, and 14,  which cover a wide range of
far-IR SED shapes, to derive empirical dust templates. The final set
of templates consists of four empirical templates and ten
templates from the SWIRE library covering a range of galaxy types
(e.g., elliptical galaxies, spiral galaxies, QSOs), see also Table~\ref{tab:fir}. The resulting
$\chi^{2}$ distribution and the best $\chi^{2}$ are derived by
considering all redshifts and all templates in the final set. Note
that the final $\chi^2$ curve shows the minimum $\chi^{2}$ for the
template set as a function of redshift and therefore is dependent on
the template set used.

Due to the varying spatial coverage of the supplementary multiwavelength data
there are some sources that lie out of the field in some photometric bands. These photometric data points are not included. In case of non-detections at certain wavelengths 3$\sigma$ upper limits are taken into account by {\tt hyperz}. The best fitting SED and $\chi^{2}$ are shown in Fig. \ref{fig:results} and Fig. \ref{fig:results2}. 11 SMGs (without known redshifts) have best fit SEDs that are empirical demonstrating the big advantage of using empirical templates.

We test our photometric redshift analysis by only considering
the far-IR emission, i.e. SPIRE and LABOCA photometry. We derive templates from the same sources
as above but without the MIPS detection and use the same ten
templates from the SWIRE library as before and fit to find the best 
redshift solution. The photometric redshifts found mostly agree with the previously determined redshifts. For one source (DKB15) the MIPS detection,
however, is crucial and we 
fail to find the same redshift. Our exercise shows that if the
far-IR peak and its Rayeigh-Jeans tail are well sampled by
observations, we can construct reliable photometric redshifts from
these data alone with typical uncertainties of ~30\% that allow us to conclude if a source can belong to a structure associated with MRC1138 or not. If this is not the case then shorter wavelength data
are crucial for constraining photometric redshifts. We note that redshifted [CII] emission contributes to the SPIRE 500~$\mu$m flux  for sources at $z\approx2.2$, see e.g. DKB01, DKB07, DKB12b in Fig.~\ref{fig:results}. This is consistent with \citet{sey12} who describe the contribution of the [CII] emission to the SPIRE 500~$\mu$m flux  for MRC1138$-$262. See also \citet{sma11} for a detailed discussion of the effect of far-IR lines on far-IR/submm broad-band fluxes.

Fig.~\ref{fig:histo} shows the distribution of photometric
redshifts for all 16 LABOCA sources. The results suggest that a significant fraction (about 50\%) of the submillimeter sources are
consistent with being protocluster members. 

\begin{figure}
\centering \includegraphics[scale = 0.35]{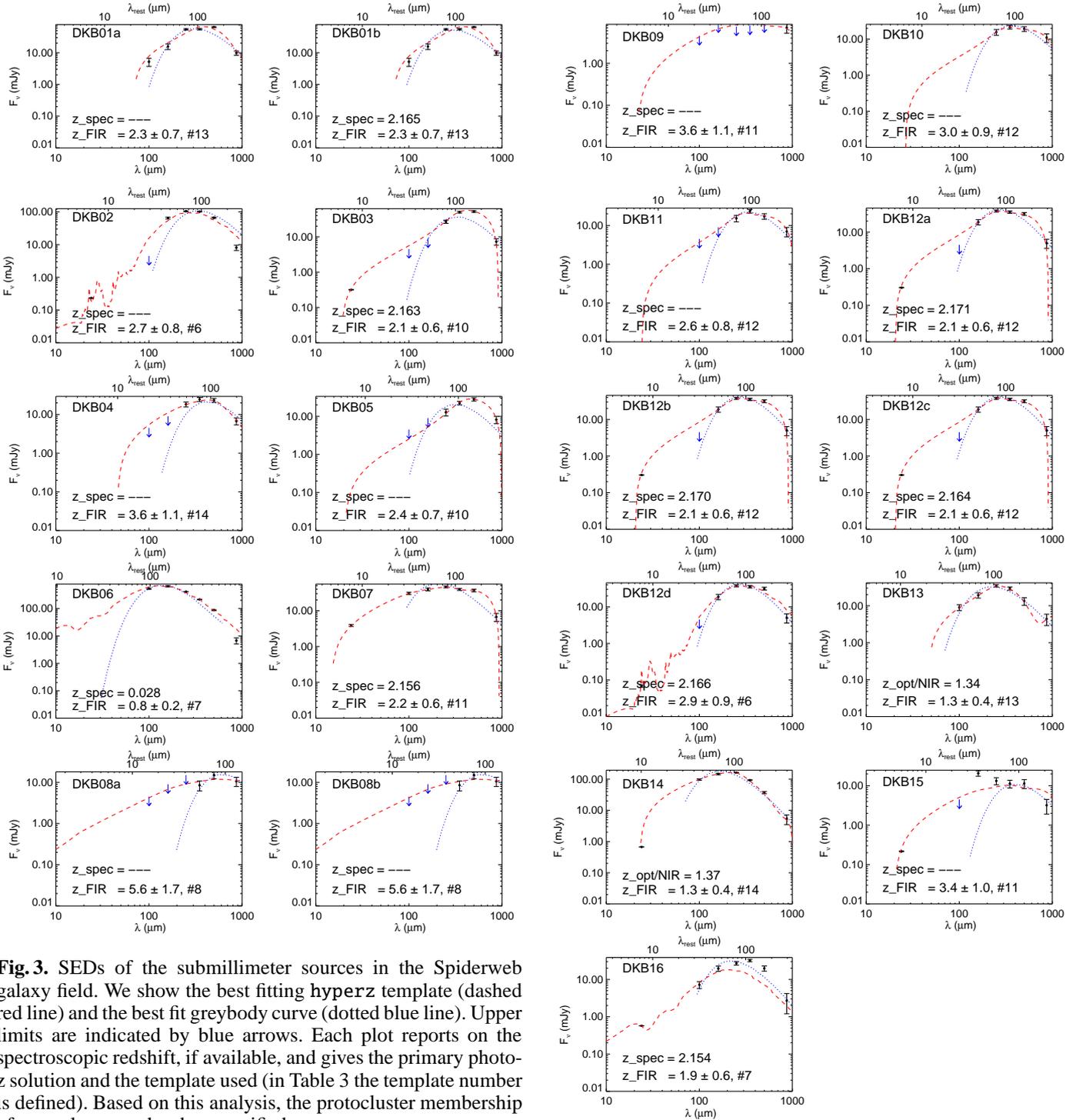}
\caption{SEDs of the submillimeter sources in the Spiderweb galaxy field. We
  show the best fitting {\tt hyperz} template (dashed red line) and
  the best fit greybody curve (dotted blue line). Upper limits are
  indicated by blue arrows. Each plot reports on the spectroscopic
  redshift, if available, and gives the primary photo-z
  solution and the template used (in Table~\ref{tab:fir} the template number is defined). Based on this analysis, the protocluster membership of
  several sources has been verified.}
\label{fig:results}
\end{figure}
\setcounter{figure}{2}
\begin{figure}
\centering \includegraphics[scale = 0.35]{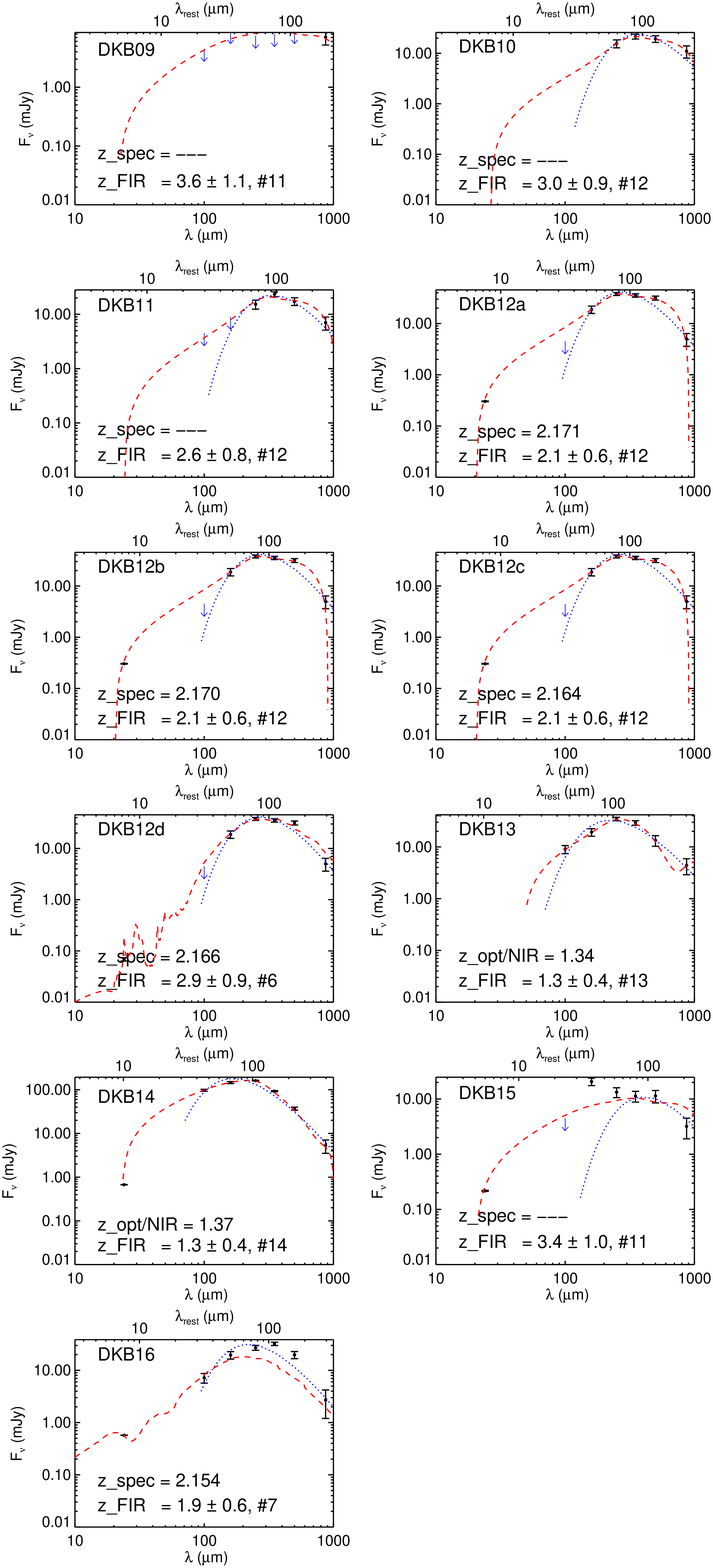}
  \caption{(cont'd)}
 \label{fig:results}\end{figure}

\begin{figure}
\centering \includegraphics[scale = 0.35]{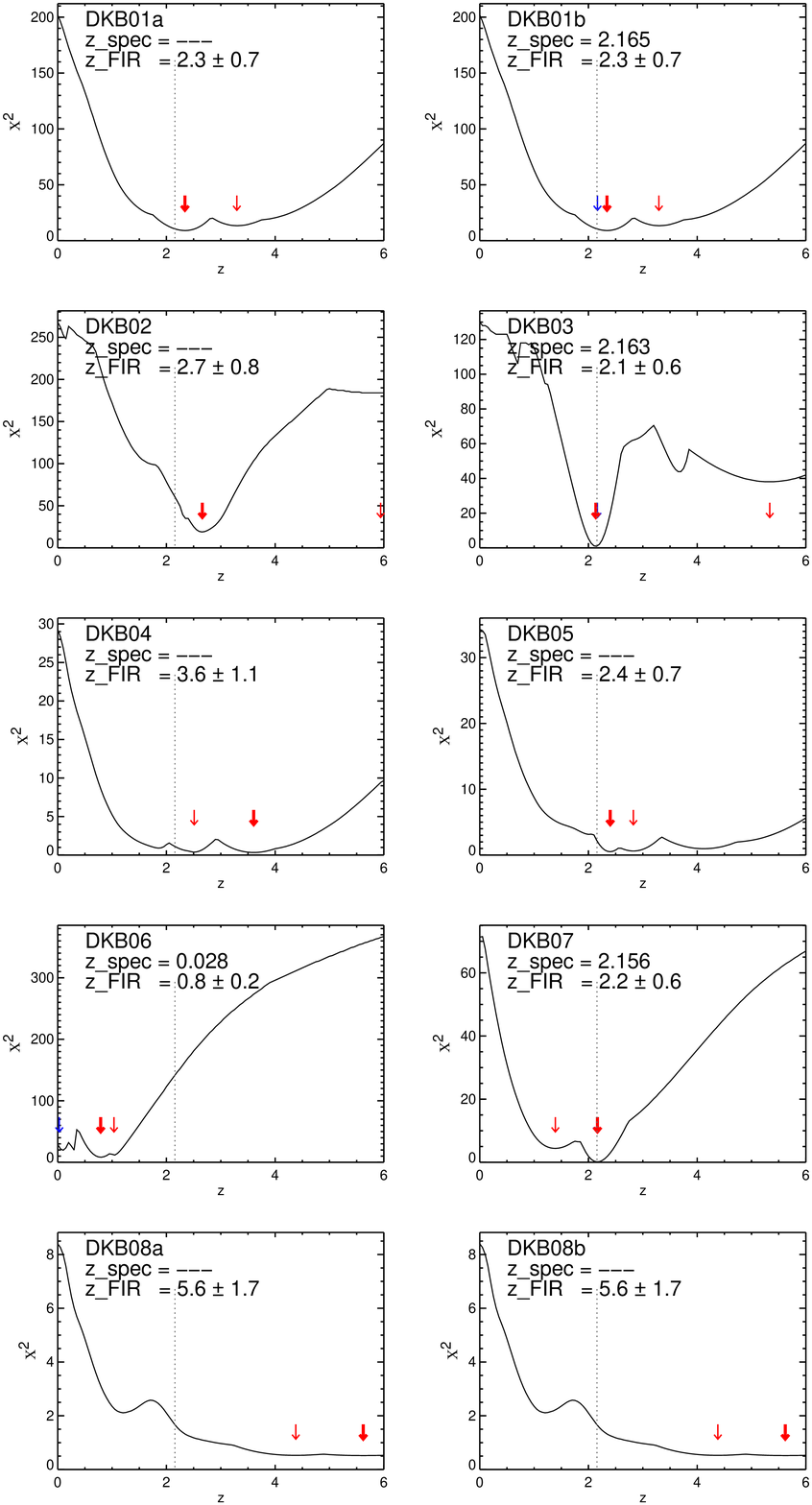}
\caption{Minimum $\chi^{2}$ as a function of redshift for the submillimeter
  sources in the Spiderweb galaxy field. Thick red arrows show the
  redshift of the primary solution of the photo-z fitting, thin red
  arrows show the redshift of the secondary solution and blue arrows spectroscopic redshifts. The grey dashed
  line indicates the redshift of the Spiderweb galaxy.}
\label{fig:results2}
\end{figure}
\setcounter{figure}{3}
\begin{figure}
\centering \includegraphics[scale = 0.35]{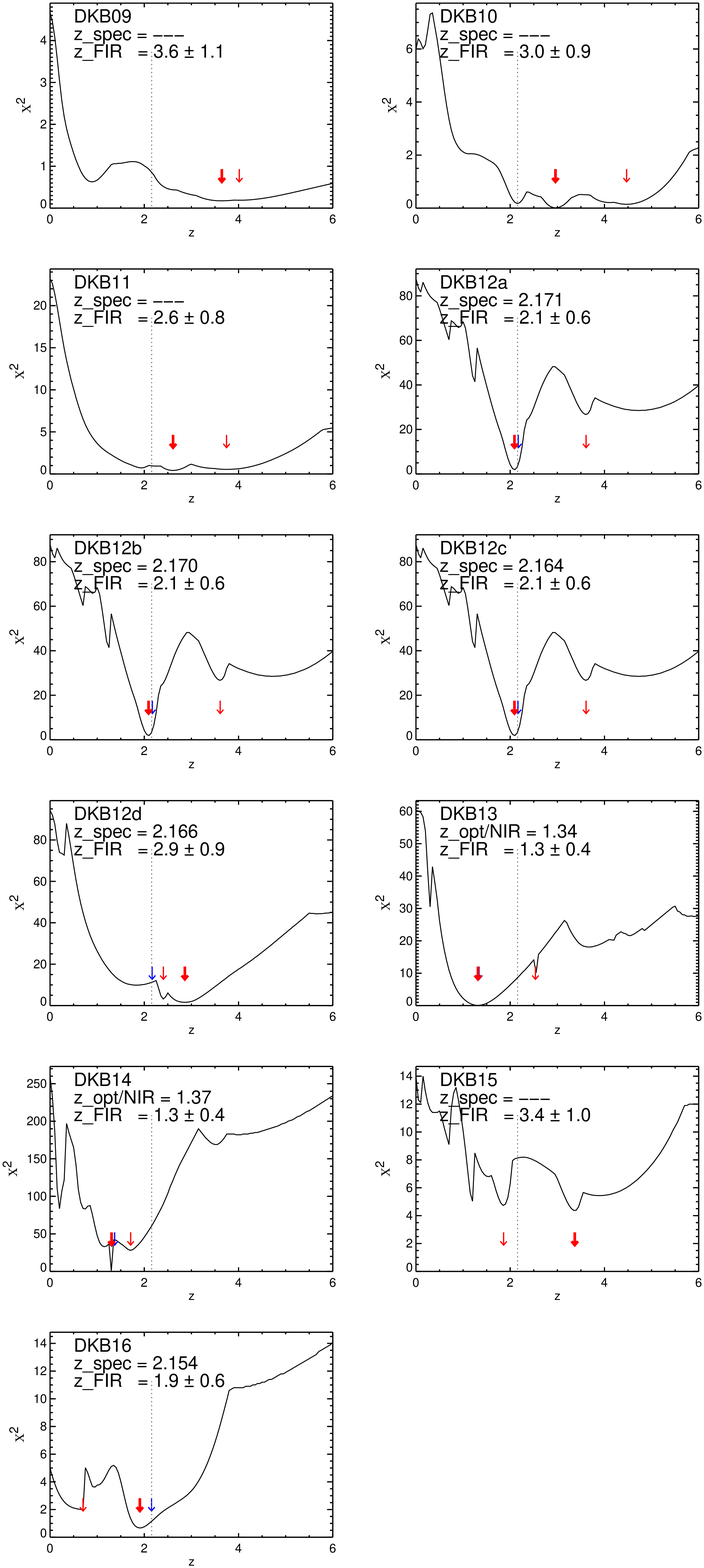}
  \caption{(cont'd)}
\end{figure}

\begin{figure}
\centering \includegraphics[scale = .19]{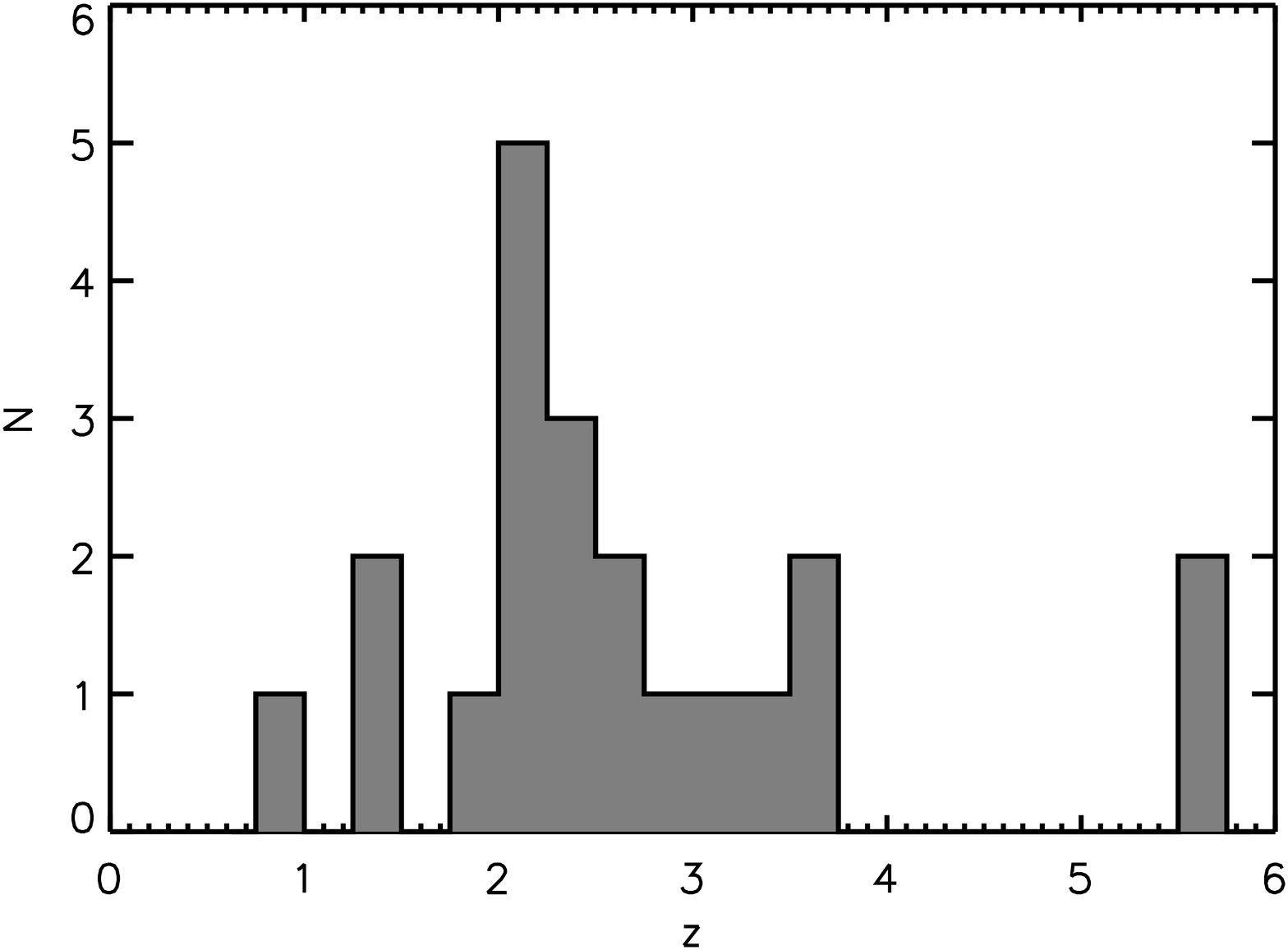}
\caption{Histogram of the far-infrared photometric redshifts (grey
  filled histogram). Our
  far-IR photometric analysis suggests that a significant fraction of the LABOCA sources
  are associated with the protocluster around the Spiderweb galaxy at
  $z=2.2$.}
\label{fig:histo}
\end{figure}

For sources with four or more detections in the {\it Herschel} and
LABOCA bands we derive dust temperatures, far-IR luminosities and star
formation rates (SFR), see Table~\ref{tab:fir}. These sources were fit
with a grey-body law of the form: $S_{\nu} \propto
\nu^{\beta}B_{\nu}(T_{d}) = \frac{\nu^{\beta+3}}{(e^{h\nu/kT_{d}}-1)}$
where $S_{\nu}$ is the flux density at the rest-frame frequency $\nu$,
$\beta$ the grain emissivity index and $T_{d}$ the dust
temperature. Dust temperatures for an interstellar medium only heated
by star formation are expected to range between $\sim$ 20--60 K, and
$\beta$ can range between 1--2.5 \citep{cas12}. For some sources spectroscopic redshift are known (Table~\ref{tab:fir}) and we use them for the conversion to rest-frame flux density from which $T_{d}$ and $\beta$ are inferred through fitting. If no spectroscopic information is available, we use the photometric redshifts but also fix $T_{d}$ to 35~K and $\beta$ to 1.5. This allows us to overcome the well known $T_{d}-z$ degeneracy \citep{bla02}. If only three detections in the far-IR are available, we also fix $T_{d}$ and $\beta$ and are thus able to estimate $L_{far-IR}$ and SFR. Far-IR luminosities
are derived by integrating the SED over the wavelength range
$8-1000$~$\mu$m and applying the relation $L_{\rm{FIR}} = 4\pi
D_{L}^{2}F_{\rm{FIR}}$ where $D_{L}$ is the luminosity distance
computed from their photometric redshifts (spectroscopic redshifts are
used if they are available). We then estimate the star formation rates
by using $SFR \ [M_{\astrosun}] = L_{\rm{FIR}}/5.8\times 10^{9}$
L$_{\astrosun}$ \citep{ken98} \footnote{We note that if we used the Chabier IMF \citep{cha03} the SFRs would decrease by a factor of 1.8.}. The derived SFR of
1800~M$_{\odot}$~yr$^{-1}$ for MRC1138$-$262 (DKB7) agrees well with
the SFR found by \citet{sey12}.


\section{Notes on individual objects}
\label{sec:notes}
In this section we discuss secure and possible counterparts (based on the p-statistic) for
all 16~LABOCA sources and whether the source could be a member of the
large scale structure at redshift $z=2.16$. SMG counterparts with spectroscopical confirmation at the redshift of the protocluster and consistent far-IR photometric redshift are classified as secure members of this large scale structure at $z\approx2.2$. We assess a protocluster membership as possible if the optical/near-infrared photometric redshift respectively the selection as an HAE without spectroscopic confirmation is strengthened by far-IR photo-z. We exclude the membership if optical/near-infrared and far-IR photo-z are discrepant. Remaining sources where no reliable judgement on the cluster membership can be made, we classify as uncertain members.

For each LABOCA
source we show 40$^{\prime\prime}$$\times$40$^{\prime\prime}$ images
at radio, far-IR, 24$\mu$m and H$\alpha$ wavelengths (Fig.~\ref{fig:cutouts}).  We only discuss
sources within the 9.5$^{\prime\prime}$ search radius (the FWHM of
the LABOCA beam). At redshift $z=2.16$, 1~arcsec corresponds to
8.4~kpc.\\

'Secure' LABOCA sources
\newline

{\it SMM~J114100.0$-$263039 (DKB01, protocluster member YES)} --- There are two
HAEs located $4\farcs5$ and $5\farcs7$ from the
nominal LABOCA position. Both HAEs are classified as secure counterparts
by the p-statistic and are only separated by 4$\farcs5$ or 37.8~kpc. This counterpart has a spectroscopic
redshift of 2.165 (Kurk et al. in prep.). The far-infrared photo-z
$z_{FIR}=2.3\pm0.7$ is consistent with the spectroscopic redshift. The
association of {\it Herschel} flux with DKB01b strongly suggests that this
source is (partly) responsible for the submillimeter emission. The closer HAE (DKB01a) has an optical/near-infrared photometric redshift of
$z_{opt/NIR}=2.84^{+0.05}_{-0.03}$ \citep{tan10} and the redshift has been very recently confirmed to be $z=2.2$ \citep{shi14}.  The {\it Herschel} detection appears to be associated to the
more distant HAE, DKB01b. DKB01 and DKB03 are only separated by 
 $24\farcs4$  (205~kpc) and the LABOCA map shown in Fig.~\ref{fig:labocasignal_noise} indicates that these two sources
could be connected to each other.\\

{\it SMM~J114053.3$-$262913 (DKB02, (P)OSSIBLE MEMBER)} --- Within the LABOCA
beam we find a prominent X-ray source: an AGN at $z=1.512$
\citep[X9][]{pen02,cro05}. However, the secure VLA association
($5\farcs4$ away) is not associated with this X-ray source. The location of the {\it Herschel} PACS counterpart suggests that the
LABOCA emission is related to the VLA source, $2\farcs4$
 from the nominal LABOCA position, and not to the AGN at $z=1.5$.  No spectroscopic redshift is known for this radio source, but both the
optical/near-infrared photometric redshift of $z_{opt/NIR}=2.12^{+0.13}_{-0.20}$ \citep{tan10}
and the far-IR photometric redshift of $z_{FIR}=2.7\pm0.8$ means this SMG is a plausible 
protocluster member. Finally, we note
that $8\farcs2$  away from the nominal bolometric
position lies an HAE candidate from the sample of \citet[\# 211
  in][]{kur04a}. Due to their shallower H$\alpha$ data
\citet{koy13a} do not recover this source. This source has no 
spectroscopic confirmation and lies $10\farcs4$ from the VLA
position.\\

{\it SMM~J114058.3$-$263044 (DKB03, YES) } ---  $7\farcs1$ away from the nominal
LABOCA position we find a spectroscopically confirmed HAE
($z_{spec}=2.163$; Kurk et al. in prep.) which is classified as secure
by the corrected Poissonian probability. Neither VLA nor PACS
counterparts are associated with this LABOCA source. However, DKB03
is detected at all three SPIRE wavelengths and the
$z_{FIR}=2.1\pm0.7$ is consistent with the spectroscopic
redshift of the H$\alpha$ counterpart. The 250~$\mu$m source position lies $9\farcs5$
from the nominal LABOCA position, but only $3\farcs4$ from the
confirmed H$\alpha$ emitter, suggesting that this near-infrared excess
source emits (some of) the dust emission detected by LABOCA. \\

{\it SMM~J114046.8$-$262539 (DKB04, (U)NKNOWN) } --- This
source is only covered by the {\it Herschel} and VLA imaging. A faint, secure,
$3.3\sigma$ VLA counterpart, not detected by {\it Herschel}, lies $7\farcs0$ from the
nominal LABOCA position.\\

{\it SMM~J114043.9$-$262340 (DKB05, U)} --- Only
{\it Herschel} and VLA coverage exist of this $4.5\sigma$ LABOCA
detection. A {\it Herschel} source detected at all three SPIRE bands is associated with this SMG.\\

{\it SMM~J114059.5$-$263200 (DKB06, NO)} --- At the position of this
SMG, we find a local spiral galaxy at $z=0.028$ \citep{jon09}. A
121.7~$\mu$Jy faint radio source lies $5\farcs9$ from the nominal
bolometer position. This VLA counterpart is detected by the Chandra
X-ray Observatory \citep[X14 in][]{pen02}. Within the LABOCA beam two
more X-ray sources are found by the same authors (X13 and
X15). According to \citet{pen02} all three sources are related to the
spiral galaxy. The spiral galaxy is also detected at two IRAS bands,
at 60 and 100~$\mu$m \citep[IRAS~F11384$-$2615;][]{mos90}. Based on
the IRAS colours \citep{per87} we estimate an infrared luminosity
$L_{IR}=4.1\times10^{8} L_{\astrosun}$. The PACS flux measured at
100~$\mu$m of 530.9$\pm$26.6~mJy is lower than the IRAS flux of
790$\pm$180~mJy at the same wavelength. The VLA counterpart is
detected at PACS wavelengths as well. This source is the brightest
object in the {\it Herschel} images. As far as we know this object is
one of the lowest redshift SMGs discovered by blind submillimeter
ground based surveys. Only a handful of SMGs in the local Universe are
known \citep[e.g.,][]{web03a,cha05}. However the submillimeter source
may lie behind the spiral galaxy. The far-infrared photometric redshift discussed in
Section \ref{sec:firprop} suggests $z_{FIR}=0.8\pm0.2$, however, the
$\chi^{2}$ distribution shown in Fig.~\ref{fig:results2} indicates lower
redshift solutions are also plausible.  PACS would not be able to
detect the very low infrared luminosity of $L_{IR}=8.6\times10^{9} L_{\astrosun}$ if
it was emitted at $z\approx1$. Furthermore, due the sensitivity of the
IRAS satellite, the IRAS flux can only be associated to the spiral
galaxy. .\\

{\it SMM~J114048.4$-$262914 (DKB07, YES)} --- \citet{sey12} discuss in detail 
the infrared properties of this radio galaxy, MRC1138$-$262. The LABOCA
detection seems to be slightly elongated, which is also seen in the SPIRE bands
at 350 and 500~$\mu$m \citep{sey12} and in SCUBA 850~$\mu$m data \citep{ste03}, see also section \ref{sec:scuba} for more details. This feature may be well due to multiple sources blended together, see also e.g., \citet{kar13} and \citet{hod13} for details on this topic.\\

{\it SMM~J114033.9$-$263125 (DKB08, U)} --- Within the LABOCA beam we
find two promising counterparts. The secure VLA source (DKB08b; $S_{1.4~GHz}=70.9\pm19.0~\mu$Jy) lies
at the edge of the LABOCA beam ($8\farcs5$ from the nominal
bolometer position). At a distance of $5\farcs4$ from the
LABOCA position, lies a bright {\it Herschel} H$\alpha$ emitter. Unfortunately, no optical/NIR photo-z exists for these
sources.\\

{\it SMM~J114040.9$-$262555 (DKB09, U)} --- This LABOCA source is undetected at VLA and {\it Herschel} wavelengths and is not covered by H$\alpha$ imaging. \\

{\it SMM~J114043.7$-$262216 (DKB10, U)} --- This is our brightest
LABOCA source ($S_{870~\mu m}=11.0\pm3.0$~mJy), located at the edge and
thus at the noisiest part of our LABOCA map. It is the only SMG
without PACS coverage. The reliability of the LABOCA detection is strengthened by a SPIRE source which is $8\farcs1$ separated from the nominal LABOCA position and peaks at 350~$\mu$m.\\

{\it SMM~J114038.5$-$263201 (DKB11, U)} --- At the edge of the LABOCA
beam, $8\farcs0$ away from the nominal LABOCA position, we
find a 350~$\mu$m peaker which could lie at $z=2.6\pm0.7$. Only
$4\farcs3$ from the LABOCA position lies a candidate
Ly$\alpha$ emitter \citep[\#73 in][]{kur04a}. However, \citet{cro05}
reveal a spectroscopic redshift of 0.671 for this source.\\

{\it SMM~J114057.6$-$262933 (DKB12, YES)} --- This $3.6\sigma$ LABOCA
detection is the most complex source in our sample. A 162.7~$\mu$Jy
bright 20~cm source lies $3\farcs2$ away from the nominal LABOCA
position. Four HAEs, separated by only $5\farcs5$
(46.2~kpc) lie within the LABOCA beam. Two emitters show a strong
H$\alpha$ line at $z=2.170$ (DKB12a) and $z=2.163$ (DKB12c) in ISAAC
spectroscopy (Kurk et al. in prep.).  SINFONI 3D spectroscopy of this
complex source confirms that the components 12b (VLA counterpart) and
12d lie at similar redshifts. From the SINFONI observations we obtain
following spectroscopic redshifts: 12a: $z=2.171$, 12b: $z=2.170$,
12c: $z=2.164$, and 12d: $z=2.166$ (Kurk et al., in prep.). The latter
source is also selected as a Ly$\alpha$ emitting candidate
\citep{kur04a}. The PACS beam has a FWHM of $6\farcs0$ at 100~$\mu$m
so it is impossible to associate the {\it Herschel} fluxes to one or more
of these components directly. Millimeter interferometric observations
are crucial in order to reveal the locations of the dust emission
within this complex.\\

'Cross-identified tentative' LABOCA sources
\newline

{\it SMM~J114048.3$-$262748 (DKB13, NO)} --- We find a secure 20~cm
radio source $6\farcs0$ away from the nominal LABOCA position. The
radio source seems to be associated with {\it Herschel} detections at all
five bands. However, the {\it Herschel} SPIRE colours exclude protocluster
membership and favour a lower redshift. This finding is consistent
with the derived optical/NIR photometric redshift of $z_{opt/NIR}=1.34^{+0.10}_{-0.07}$
\citep{tan10} for the VLA source.\\

{\it SMM~J114042.4$-$262715 (DKB14, NO)} --- Only
$1\farcs8$ away from the nominal LABOCA position lies the brightest radio counterpart
(618.5~$\mu$Jy) of an SMG in our sample. This VLA source
is detected by {\it Herschel} at all five bands. Again the {\it Herschel}
colours imply a low redshift. The optical-NIR counterpart
photometric redshift suggests $z_{opt/NIR}=1.37^{+0.08}_{-0.07}$
\citep{tan10}. Both redshift estimates exclude DKB14 from being a 
protocluster member.\\

{\it SMM~J114054.3$-$262800 (DKB15, P)} --- $7\farcs4$ away from the
nominal LABOCA position we find a tentative HAE association with  $z_{opt/NIR}=2.60^{+0.24}_{-0.24}$. We detect two
PACS~160~$\mu$m counterparts (one classified as secure and the other is not secure)
that are separated by only $11\farcs2$, which are undetected at PACS 100~$\mu$m.  The secure PACS~160~$\mu$m counterpart seems to be physically associated with the HAE. No other SMG in our sample has two PACS counterparts and no such system was seen in
GOODS-N  \citep{dan10}. At the edge of the PACS 160~$\mu$m
beams, we find the candidate Ly$\alpha$ emitters L877
\citep{kur04a}. However, this LAE lies at $z_{spec}=0.863$
\citep{cro05}. The far-IR photometric analysis (Fig.~\ref{fig:results}) does not exclude $z=2.2$ as a possible far-IR photometric redshift for the HAE.\\

{\it SMM~J114102.7$-$262746 (DKB16, YES) } --- This source has a wide multiwavelength coverage. \citet{pen02} report X-ray emission
for this source (X16), \citet{kur04a} selected this source as LAE
candidate (L778) and it is detected at 1.4~GHz. Subsequent rest-frame UV-spectroscopy by
\citet{cro05} reveal both the redshift $z_{spec}=2.149$ and the AGN
nature of this source. An H$\alpha$ line was detected at $z=2.154$ by Kurk et al. (in prep) and the width of the H$\alpha$ line is consistent
with the AGN nature of this source. The velocity offset between the Ly$\alpha$ and
H$\alpha$ line is $+$476~km~s$^{-1}$ which is typical for LAEs and
LBGs \citep{sha03} indicating gas outflow from this source.\\

\setcounter{figure}{5}
\begin{figure*}[!]
 \centering
\includegraphics[width=18cm,angle=0]{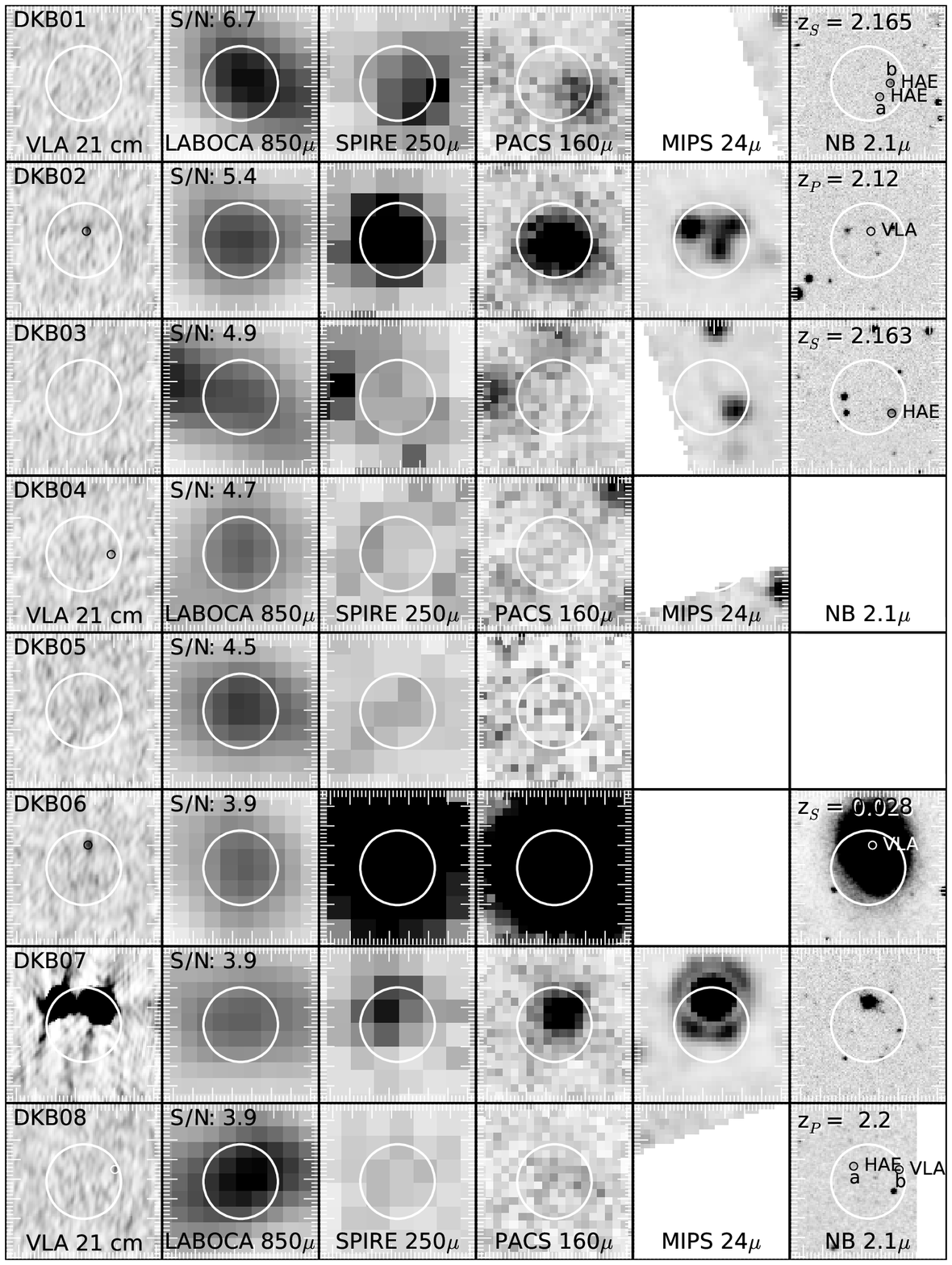}
 \caption{Postage stamps of all 16 LABOCA sources, including
   VLA~1.4~GHz, LABOCA 870~$\mu$m, SPIRE~250~$\mu m$, PACS~160~$\mu
   m$, MIPS~24~$\mu m$ and MOIRCS $z=2.16$ H$\alpha$ images. The
   40$^{\prime\prime}$$\times$40$^{\prime\prime}$ images are centered
   on the nominal LABOCA position and orientated such that north is at
   the top and east is to the left.  The large white circle represents
   the size of the LABOCA beam ($\sim$19$^{\prime\prime}$
   diameter). Small circles are VLA and/or HAE sources. Spectroscopic (S) and photometric (P) redshifts are labeled in the top of the H$\alpha$ images.}
 \label{fig:cutouts}
\end{figure*}

\setcounter{figure}{5}
\begin{figure*}[!]
 \centering
 \includegraphics[width=18cm,angle=0]{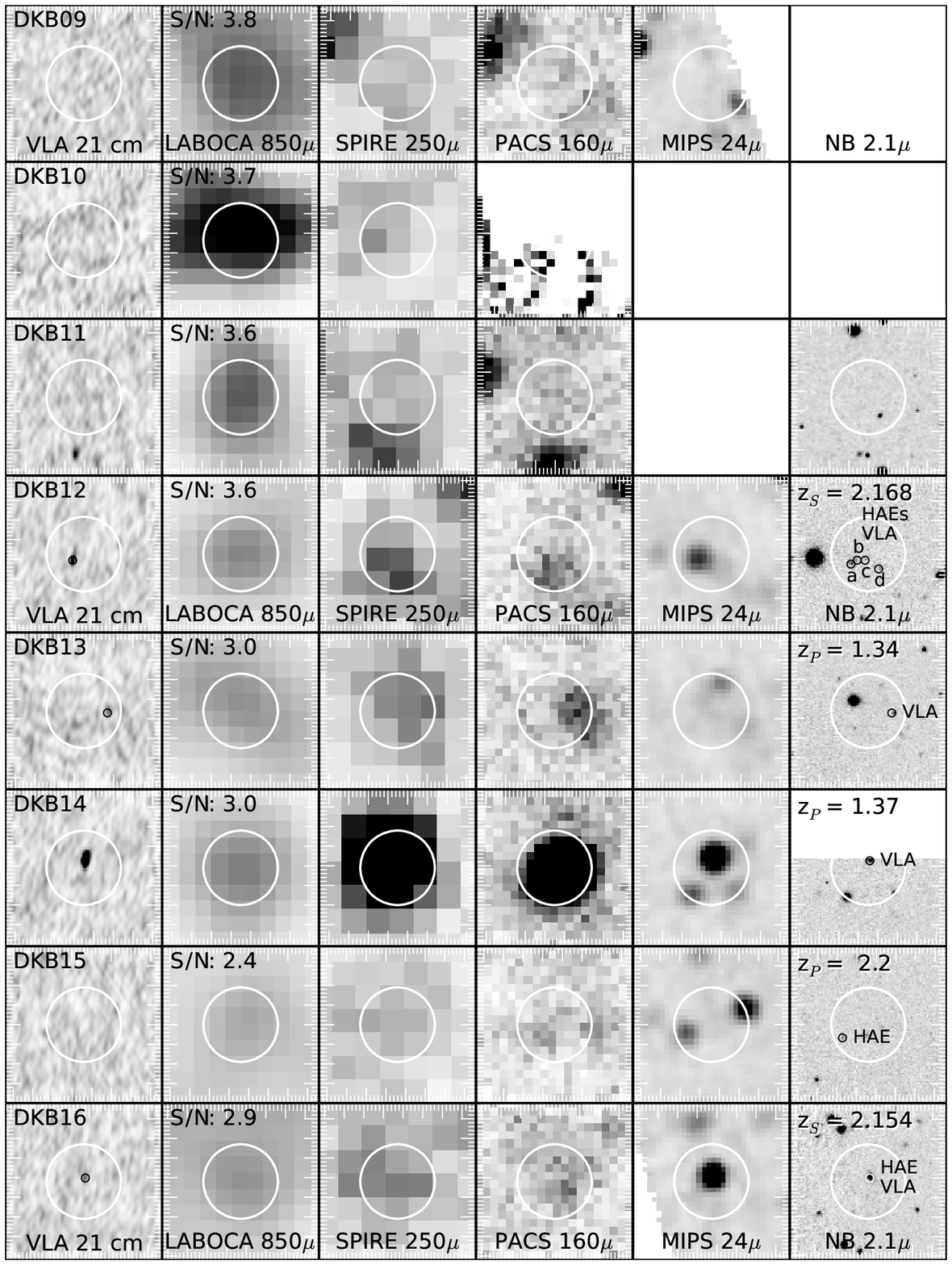}
  \caption{(cont'd)}
 \label{fig:cutouts}
\end{figure*}


\begin{landscape}
\begin{table}
\scriptsize
\caption{Main properties of counterparts of the 870~$\mu$m LABOCA sources in the field arround MRC1138$-$256}
\label{tab:id}
\begin{tabular}{ccccccccccccccc}

Source&Alias&R.A.&Decl.&Separation&z$_{opt/NIR}$&$S_{24~\mu m}$&$S_{100~\mu m}$&$S_{160~\mu m}$&$S_{250~\mu m}$&$S_{350~\mu m}$&$S_{500~\mu m}$&$S_{870~\mu m}$&$S_{1.4~GHz}$&Counterpart\\
(IAU)&&(J2000)&(J2000)&(arcsec)&&($\mu$Jy)&(mJy)&(mJy)&(mJy)&(mJy)&(mJy)&(mJy)&($\mu$Jy)&\\
(1)&(2)&(3)&(4)&(5)&(6)&(7)&(8)&(9)&(10)&(11)&(12)&(13)&(14)&(15)\\
\hline\hline
SMM~114100.0$-$263039 & DKB01a &11:40:59.82 &$-$26:30:42.6& 4.5 & 2.84$^{+0.05}_{-0.03}$ & ... & $5.2\pm1.5$  & $15.7\pm3.1$ & $54.4\pm2.7$ & $55.2\pm2.7$ & $64.5\pm2.6$ & $9.8\pm1.5$ & $<57.0$ & {\bf HAE} \\
SMM~114100.0$-$263039 & DKB01b &11:40:59.62 &$-$26:30:39.1& 5.7 &$2.165^{\star}$ & ... & $5.2\pm1.5$  & $15.7\pm3.1$ & $54.4\pm2.7$ & $55.2\pm2.7$ & $64.5\pm2.6$ & $9.8\pm1.5$ & $<57.0$ & {\bf HAE} \\
SMM~114053.3$-$262913 & DKB02 & 11:40:53.23 & $-$26:29:11.7 & 2.4 & $2.12^{+0.13}_{-0.20}$ & $232.6\pm5.0$ & $<4.5$ & $65.0\pm4.4$ & $104.4\pm3.1$ & $101.9\pm2.7$ & $66.4\pm3.0$ & $8.1\pm1.5$ & $123.2\pm19.0$ & {\bf VLA} \\
SMM~114058.3$-$263044 & DKB03 & 11:40:57.81 & $-$26:30:48.1 & 7.2 &$2.163^{\star}$ & 316.9$\pm$5.0 & $<4.5$       & $<9.0$       & $26.8\pm2.8$ & $49.9\pm2.7$ & $52.1\pm2.8$ & $7.3\pm1.5$ & $<57.0$ & {\bf HAE} \\ 
SMM~114046.8$-$262539 & DKB04 & 11:40:46.23 & $-$26:25:39.3 & 7.0 & ... & ...  & $<4.5$  & $<9.0$ & $18.4\pm2.7$ & $24.6\pm3.0$ & $23.4\pm2.7$  & $6.8\pm1.4$ & $63.0\pm19.0$ & {\bf VLA} \\
SMM~114043.9$-$262340 & DKB05 & &&& ... & ... & $<4.5$       & $<9.0$       & $12.8\pm2.6$ & $22.4\pm2.3$ & $27.9\pm2.7$ & $8.2\pm1.8$ & $<57.0$ & \\
SMM~114059.5$-$263200 & DKB06 & 11:40:59.46 & $-$26:31:54.9 & 5.9 &$0.028^{\diamond}$ & ... & $530.9\pm26.6$& $652.0\pm32.7$& $413.8\pm3.1$& $214.9\pm2.7$ & $87.7\pm3.1$ & $6.8\pm1.7$  & $121.7\pm19.0$ & {\bf VLA} \\
SMM~114048.4$-$262914 & DKB07 & &&& $2.160^{\dagger}$ & $3890.0\pm250.0$ & $30.7\pm2.2$ & $42.3\pm3.0$ & $46.1\pm2.8$ & $39.5\pm2.6$ & $36.8\pm3.2$ & $6.7\pm1.7$ & $8710\pm35$ & HzRG ({\bf HAE}, 2$\times$HAE)\\
SMM~114033.9$-$263125 & DKB08a & 11:40:34.16 & $-$26:31:21.7 & 5.4 & $2.2$ & ... & $<4.5$ & $<9.0$ & $<15.0$  & $8.5\pm2.3$ & $15.0\pm2.8$  & $10.6\pm2.7$ & $<57.0$ & HAE \\
SMM~114033.9$-$263125 & DKB08b & 11:40:33.29 & $-$26:31:22.5 & 8.5 & ... & ... & $<4.5$ & $<9.0$ & $<15.0$  & $8.5\pm2.3$ & $15.0\pm2.8$  & $10.6\pm2.7$ & $70.9\pm19.0$ & {\bf VLA } \\
SMM~114040.9$-$262555 & DKB09 & &&& ... & ... & $<4.5$ & $<9.0$ & $<7.5$ & $<8.0$ & $<9.0$  & $7.1\pm1.9$ & $<57.0$ & \\
SMM~114043.7$-$262216 & DKB10 & &&& ... & ... & ... & ... & $15.6\pm2.8$ & $21.5\pm2.4$ & $19.3\pm2.9$ & $11.0\pm3.0$ & $<57.0$ & \\
SMM~114038.5$-$263201 & DKB11 & &&& ... & ... & $<4.5$ & $<9.0$ & $15.3\pm2.9$ & $23.4\pm2.5$ & $17.3\pm2.8$ & $7.0\pm1.9$ & $<57.0$ & \\
SMM~114057.6$-$262933 & DKB12a & 11:40:57.91 & $-$26:29:36.3 & 5.2 & $2.171^{\star}$ & $303.4\pm5.0$ & $<4.5$ & $18.8\pm3.1$ & $37.8\pm2.5$ & $35.4\pm2.7$ & $31.5\pm2.8$ & $5.0\pm1.4$  & $<57.0$  & {\bf HAE} \\
SMM~114057.6$-$262933 & DKB12b & 11:40:57.79 & $-$26:29:35.3 & 3.2 & $2.170^{\star}$ & $303.4\pm5.0$ & $<4.5$ & $18.8\pm3.1$ & $37.8\pm2.5$ & $35.4\pm2.7$ & $31.5\pm2.8$ & $5.0\pm1.4$ & $162.7\pm19.0$ & {\bf VLA$^{\ast}$, HAE} \\
SMM~114057.6$-$262933 & DKB12c & 11:40:57.64 & $-$26:29:35.3 & 1.8 & $2.164^{\star}$ & $303.4\pm5.0$ & $<4.5$ & $18.8\pm3.1$ & $37.8\pm2.5$ & $35.4\pm2.7$ & $31.5\pm2.8$ & $5.0\pm1.4$ & $<57.0$ & {\bf HAE} \\
SMM~114057.6$-$262933 & DKB12d & 11:40:57.38 & $-$26:29:37.5 & 4.7 & $2.166^{\star}$ & $68.5\pm5.0$ & $<4.5$ & $18.8\pm3.1$ & $37.8\pm2.5$ & $35.4\pm2.7$ & $31.5\pm2.8$ & $5.0\pm1.4$ & $<57.0$ & {\bf HAE} \\
SMM~114048.3$-$262748 & DKB13 & 11:40:47.89 & $-26$:27:48.5 & 6.1 & $1.34^{\bullet}$ & $<15.0$ & $9.0\pm1.6$ & $19.3\pm3.2$ & $34.8\pm2.7$ & $29.2\pm2.9$ & $13.4\pm3.1$ & $4.4\pm1.5$ & $93.7\pm19.0$ & {\bf VLA} \\ 
SMM~114042.4$-$262715 & DKB14 & 11:40:42.35 & $-$26:27:13.7 & 1.8 & $1.37^{\bullet}$ & $672.4\pm5.0$ & $97.4\pm5.1$ & $146.1\pm7.9$& $161.0\pm2.9$& $92.7\pm2.6$ & $37.1\pm3.0$& $5.3\pm1.8$ & $618.5\pm19.0$ & {\bf VLA} \\
SMM~114054.5$-$262800 & DKB15 & 11:40:54.75 & $-$26:28:03.4 & 7.4 & $2.2$ & $216.7\pm5.0$ & $<4.5$  & $20.7\pm3.2$  & $13.3\pm2.7$ & $11.3\pm2.5$ & $11.4\pm2.9$ & $3.2\pm1.3$ & $<57.0$ & HAE \\
SMM~114102.7$-$262746 & DKB16 & 11:41:02.38 & $-$26:27:45.1 & 1.0 & $2.154^{\star\ddag}$ & $572.1\pm5.0$ & $7.2\pm1.5$ & $19.6\pm3.2$  & $27.1\pm2.9$ & $32.2\pm2.4$ & $19.8\pm3.1$ & $4.2\pm1.4$ & $76.2\pm19.0$ & {\bf HAE$^{\ast\ast}$, VLA} \\
\hline
\end{tabular}
\tablefoot{--- Col.~(1): LABOCA source. Col.~(2): Short name of
  LABOCA source. Col.~(3-4): J2000.0 coordinates of associated LABOCA
  counterparts either from VLA or H$\alpha$
  observations.  Col.~(5): Separation between nominal LABOCA bolometer and
  counterpart position. Col.~(6): Spectrosopic (three digits after the
  decimal), photometric redshift (two digits after the decimal) and
  H$\alpha$ imaging (one digit after the decimal) of the LABOCA
  counterpart. $^{\star}$: Kurk et al. in prep.; $^{\diamond}$:
  \citet{jon09}; $^{\dagger}$: \citet{kui11}; $^{\bullet}$:
  \citet{tan10}. $^{\ddag}$: Previously, \citet{cro05} obtained a
  rest-frame UV-spectroscopic redshift of $z=2.149$ for the
  counterpart of DKB16. Col.~(7-14): Flux measurements with {\it Spitzer}, {\it
    Herschel}, LABOCA and VLA. For multiple component counterparts as
  DKB01, DKB08 and DKB12, we give for the individual components the
  {\it Spitzer} and {\it Herschel} fluxes associated to the LABOCA
  detection. For the {\it Herschel} bands we give source detection errors. In case of the VLA measurements, we list the peak
  flux. Col.~(15): Type of LABOCA counterpart. Secure counterparts are
  in bold face. $^{\ast}$: In case of DKB12b we list the VLA
  position. $^{\ast\ast}$: In case of DKB16 we list the HAE
  position.}
\end{table}
\end{landscape}


\begin{table*}
\centering
\caption{Redshifts, star formation rates and results of the far-IR SED fitting of the 870~$\mu$m LABOCA sources in the field around MRC1138$-$256}
\label{tab:fir}
\begin{tabular}{cccccccccc}

Alias&Member&z$_{opt/NIR}$&z$_{FIR}$&SFR$_{H\alpha}$&SFR$_{FIR}$&L$_{FIR}$&T&$\beta$&Template\\
&&&&(M$_{\odot}$~yr$^{-1}$)&(M$_{\odot}$~yr$^{-1}$)&($10^{12}$~L$_{\odot}$)&(K)&\\
(1)&(2)&(3)&(4)&(5)&(6)&(7)&(8)&(9)&(10)\\
\hline\hline
DKB01a&YES&2.84$^{+0.05}_{-0.03}$&$2.3\pm0.7$&30&1320&7.6&35&1.5&\#13 (DKB13)\\
DKB01b&YES&2.165&$2.3\pm0.7$&230&1090&6.3&37&1.3&\#13 (DKB13)\\
DKB02&P&2.12$^{+0.13}_{-0.20}$&$2.7\pm0.8$&&3080&17.9&35&1.5&\#6 (I22491)\\
DKB03&YES&2.163&$2.1\pm0.6$&290&650&3.8&31&1.5&\#10 (DKB03)\\
DKB04&U&...&$3.6\pm1.1$&&1010&5.9&35&1.5&\#14 (DKB14)\\
DKB05&U&...&$2.4\pm0.7$&&490&2.8&35&1.5&\#10 (DKB03)\\
DKB06&NO&0.028&$0.8\pm0.2$&&1&0.009&22&1.9&\#7 (Mrk231)\\
DKB07&YES&2.160&$2.2\pm0.6$&&1750&10.1&56&1.0&\#11 (DKB07)\\
DKB08a&U&2.2&$5.6\pm1.7$&$>$20$^{\star}$&1460&8.4&35&1.5&\#8 (QSO1)\\
DKB08b&U&..&$5.6\pm1.7$&&1460&8.4&35&1.5&\#8 (QSO1)\\
DKB09&U&...&$3.6\pm1.1$&&$^{\dagger}$&$^{\dagger}$&$^{\dagger}$&$^{\dagger}$&\#11 (DKB07)\\
DKB10&U&...&$3.0\pm0.9$&&810&4.7&35&1.5&\#12 (DKB12c)\\
DKB11&U&...&$2.6\pm0.8$&&620&3.6&35&1.5&\#12 (DKB12c)\\
DKB12a&YES&2.171&$2.1\pm0.6$&240&860&5.0&35&1.8&\#12 (DKB12c)\\
DKB12b&YES&2.170&$2.1\pm0.6$&160&860&5.0&35&1.8&\#12 (DKB12c)\\
DKB12c&YES&2.164&$2.1\pm0.6$&30&850&4.9&35&1.8&\#12 (DKB12c)\\
DKB12d&YES&2.166&$2.9\pm0.9$&100&850&4.9&35&1.8&\#6 (I22491)\\
DKB13&NO&1.34$^{+0.10}_{-0.07}$&$1.3\pm0.4$&&280&1.6&38&1.0&\#13 (DKB13)\\
DKB14&NO&1.37$^{+0.08}_{-0.07}$&$1.3\pm0.4$&&2020&11.7&45&1.6&\#14 (DKB14)\\
DKB15&P&2.2&$3.4\pm1.0$&90&460&2.7&35&1.5&\#11 (DKB07)\\
DKB16&YES&2.154&$1.9\pm0.6$&1140&830&4.8&48&1.3&\#7 (Mrk231)\\
\hline
\end{tabular}
\tablefoot{--- Col.~(1): Short name of LABOCA
  source. Col.~(2): Classification on membership of the $z\approx2.2$
  protocluster structure. YES=secure member; P=possible member; U=no reliable statement
  on membership could be made; NO=membership securely excluded. Col.~(3): Spectrosopcic (three digits
  after the decimal), photometric redshift (two digits after the decimal)
  and H$\alpha$ imaging (one digit after the decimal) of the LABOCA
  counterpart. Col.~(4): Redshift estimate from the far-infrared
  SED. Col.~(5): The star formation rate derived from narrow-band H$\alpha$ imaging
  \citep{koy13a} is based on the NB flux and includes corrections for [NII] contamination and dust extinction following \citet{koy13a}. $^{\star}$: Due to its faintness in the K$_{s}$-band no corrections could be applied. Col.~(6): Star formation rate
  derived from our infrared luminosities estimates and using the
  conversion from \citet{ken98}. Col.~(7): Far-infrared
  Luminosity. Col.~(8): Dust Temperature. Col.~(9): Spectral index. In
  order to overcome the well-known $T-z$ degeneracy
  \citep[e.g.,][]{bla02}, we fixed the temperature $T$ to 35~K and the
  spectral index $\beta$ to 1.5 where no spectroscopic information is
  available. The same we also do for sources with only 3 detections in the far-IR bands. Col.~(10): Template used. \\
  $^{\dagger}$: DKB09 is only detected at 870~$\mu$m, therefore no physical properties are derived.}
\end{table*}

\section{Characteristics of the LABOCA overdensity}
\subsection{Previous SCUBA observations}
\label{sec:scuba}
\citet{ste03} observed a small field of $\sim$2$^{\prime}$ diameter centered on MRC1138$-$262
with SCUBA and report four detections including the radio galaxy. They reported a higher source density than expected from blank fields (by one source). However, we only recover two of these sources with our LABOCA observations, which are at a
very similar wavelength to the SCUBA observations. The other two SCUBA sources are fainter ($S_{850~\mu m}$=3.1 and 2.2~mJy) than the 3$\sigma$ detection limit of  $\sim$4~mJy\\ at this part of our LABOCA map. The fluxes of
our two LABOCA sources, the radio galaxy and DKB02, are consistent
with the ones obtained from the SCUBA observations. 
In addition, we do not find the
proposed alignment between HzRG radio axis and bright submillimeter
companions. 

However, similar to \citet{ste03}, we find that the LABOCA emission of
MRC1138$-$262 is spatially extended. \citet{sey12} find that this
extension consists of four galaxies detected by {\it Spitzer}, two of
them are spectroscopically identified to lie at the same redshift
\citep{kur04b}, cf. with \citet{ivi08,ivi12}.
\begin{figure}[!h]
 \centering
 \includegraphics[width=9cm,angle=0]{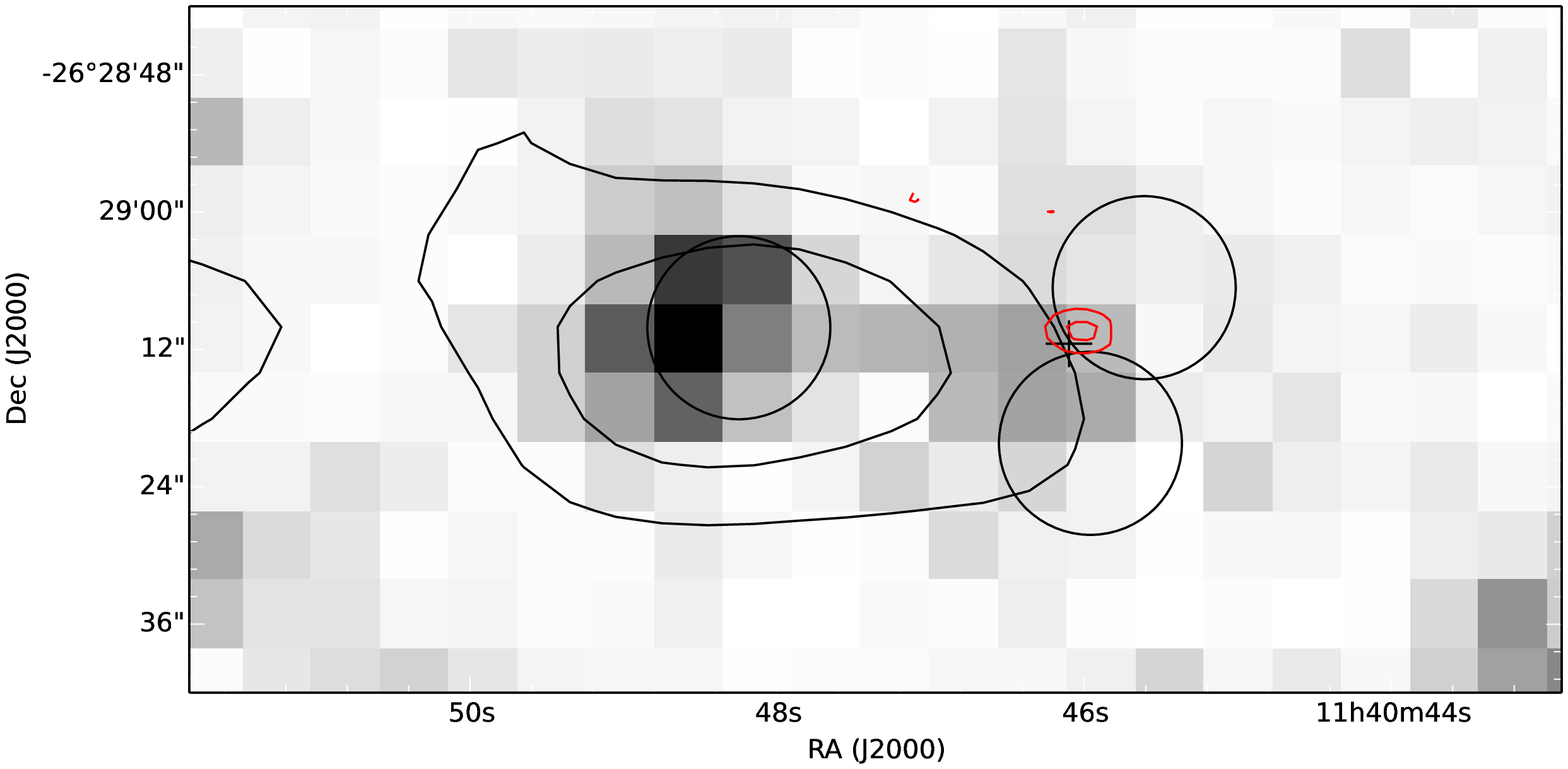}
 \caption{Environment of the radio galaxy MRC1138. SPIRE 250~$\mu$m is
   shown in greyscale. The black contours are the (extended) LABOCA
   emission of the radio galaxy, indicating levels at 2.0, 4.0 and 6.0~mJy/beam. The circles with a diameter
   of 16$^{\prime\prime}$ (the FWHM of SCUBA) indicate SCUBA emission
   \citep{ste03}.  The red contours show the location of the CO(1-0)
   emission \citep{emo13} associated to the spectroscopically
   confirmed H$\alpha$ emitter \# 229  at $z=2.15$ \citep{kur04b}. The CO levels are 0.134 and 0.168~mJy/beam.}
 \label{fig:hae229}
\end{figure}
30$^{\prime\prime}$ west of the radio galaxy, \citet{emo13} report
CO(1-0) emission at $3.7\sigma$ significance associated with the
spectroscopically confirmed H$\alpha$ emitter \#229 at $z=2.149$
\citep{kur04b} --- \citet{koy13a} also select this source as H$\alpha$ emitter. This cold molecular gas reservoir is now confirmed
through very recent ATCA observations (Emonts, priv. communication). This
CO-bright HAE lies 8.1$^{\prime\prime}$ south-east away from the
nominal SCUBA position of source \# 3 \citep{ste03}, see
Fig.~\ref{fig:hae229}. Below the published SCUBA source, we see dust
emission at low level in the SCUBA map \citep[see Fig.~1
  in][]{ste03}. SPIRE emission, coincident with the position of
HAE229, lies in between the northern and southern components and seems
to be related to the extended LABOCA emission of the HzRG. We
speculate that the northern and southern components are one submm
source but due to the chopper throw used in the SCUBA observations,
negative flux from the central HzRG was added right on the source,
cutting it into two and artificially reducing the total flux. HAE229
is detected at PACS 160~$\mu$m and at all three SPIRE bands (see
Table~\ref{tab:hae229}) and we derive a far-IR photometric redshift
$z_{FIR}=1.8\pm0.5$ assuming the total SCUBA flux of SCUBA source \#3 is
$S_{850~\mu m}=2.2\pm1.4$~mJy \citep{ste03}. However, we note that the true amount of submm emission at 850~$\mu$m is uncertain and only deeper and higher resolution observations will reveal the true configuration of this source in the submm window. We conclude that the CO-bright HAE229
is an SMG related to the protocluster at $z\approx2.2$. However, as this source is not detected by our LABOCA observations as a single source we will exclude it in the forthcoming discussion. The extended LABOCA emission of MRC1138$-$262 suggests that most probably, the LABOCA emission of HAE229 is blended with the one of the radio galaxy.
\begin{table}
\begin{center}
\caption{Fluxes of the CO-bright H$\alpha$ emitter \#229}
\label{tab:hae229}
\begin{tabular}{cccc}
Band&Flux&Unit&Instrument\\
(1)&(2)&(3)&(4)\\
\hline\hline
$S_{24~\mu m}$&$477.4\pm5.0$&$\mu$Jy&MIPS\\
$S_{100~\mu m}$&$<4.5$&mJy&PACS\\
$S_{160~\mu m}$&$13.6\pm4.0$&mJy&PACS\\
$S_{250~\mu m}$&$26.0\pm2.8$&mJy&SPIRE\\
$S_{350~\mu m}$&$27.2\pm2.9$&mJy&SPIRE\\
$S_{500~\mu m}$&$26.5\pm2.7$&mJy&SPIRE\\
$S_{850~\mu m}$&$2.2\pm1.4$&mJy&SCUBA\\
\hline
\end{tabular}
\tablefoot{--- Col.~(1): Band in which flux is measured. Col.~(2): Units of the flux density measurements. Limit of PACS~100~$\mu$m observation is $3\sigma$. Col~(3): Our measurements for HAE229. SCUBA flux is from \citet{ste03}. Col.~(4): Instruments.}
\end{center}
\end{table}

\subsection{Ly$\alpha$ emitting counterparts to LABOCA sources}
Besides MRC1138$-$262, three SMGs (DKB12, DKB15 and DKB16) are associated with
LAEs. In two cases (DKB12d and DKB16), the LAE has been confirmed by
H$\alpha$ spectroscopy (Kurk et al., in prep.). This result is in
contrast to the work by \citet{deb04} on the protocluster around the
$z=4.1$ radio galaxy TN~J1338$-$1942 who reported no associations of
confirmed LAEs with SMGs. However, this discrepancy could be explained thereby that both LAEs have been selected as HAEs  \citep{koy13a} and DKB12 is even seen at PACS
wavelengths \citep[cf.][PACS detection of 2/72 LAEs between
  $z=2.0-3.5$]{ote12}.

\subsection{H$\alpha$ emitting counterparts to LABOCA sources}
As discussed in Section~\ref{sec:hae} we find six out of 11 SMGs
covered by H$\alpha$ imaging at $z\approx2.2$ are associated with HAEs. We
search the literature for H$\alpha$ surveys of fields containing SMGs
at the redshift of the survey and find H$\alpha$ narrow
band observations of the SSA~13 field at $z\approx2.23$, which included two SMGs
at the probed redshift range \citep{mat11}. None of the two SMGs were selected as HAEs
with fluxes greater than 
$f(H\alpha)\approx1.0^{-16}$~erg~s$^{-1}$~cm$^{-2}$. The flux limit of the \citet{koy13a}
 data is  $f(H\alpha)\approx3.0^{-17}$~erg~s$^{-1}$~cm$^{-2}$. Approximately 50\% of the HAEs associated to SMGs in the MRC 1138-262 field would be missed if the  \citet{koy13a}
H$\alpha$ images were of a similar depth to the \citet{mat11} data.

All SMGs with HAEs counterparts beside one (DKB16) have a large discrepancy
between the SFR derived from H$\alpha$ and from the far-IR indicating these sources are highly 
dust-obscured \citep[consistent
  with][]{swi04}. It may demonstrate that a large amount of star
formation activity is missed when using the H$\alpha$ line as a SFR indicator \citep[see also][]{koy10}. The H$\alpha$ derived SFR (based on the
  narrow-band imaging by \citet{koy13a} and corrected for [NII] contamination and dust
  extinction following \citet{koy13a}) ranges between $\sim$30 to 300 M$_{\odot}$ yr$^{-1}$ for all beside one source (DKB16, SFR$_{H\alpha}$=1140 M$_{\odot}$ yr$^{-1}$) 
whereas the SFR derived from our IR observations ranges between 300 to
1800 M$_{\odot}$ yr$^{-1}$.

\begin{figure}[!h]
 \centering \includegraphics[width=9cm,angle=0]{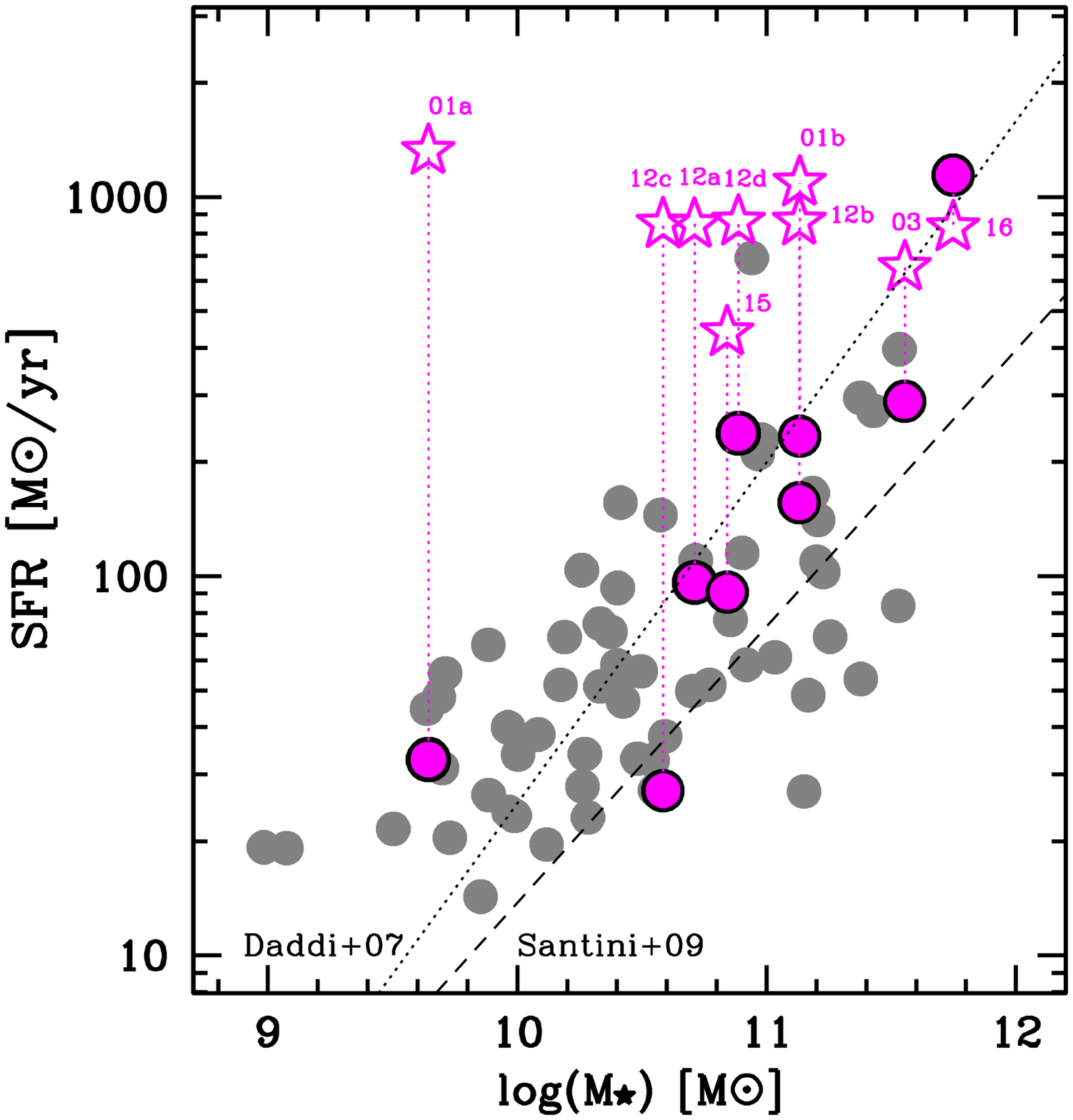}
 \caption{Stellar mass vs. star formation rate
   (derived from the H$\alpha$ line) for the complete sample of
   H$\alpha$ emitters (grey dots) discovered in the field of
   MRC1138$-$262 \citep{koy13a}. Only DKB07 and DKB08a are excluded as the estimates of the stellar mass of the associated HAEs are unreliable.
The pink dots indicate the H$\alpha$ emitters located within the LABOCA beam
   aperture for all LABOCA source. The SFR based on the far-IR measurements is shown as open pink stars (connected with dotted lines). The far-IR measurements reveal that far-IR bright HAEs (LABOCA sources) are off the star-formation
   main sequence for $z\sim2$ galaxies \citep{dad07,san09}.}
 \label{fig:massvssfr}
\end{figure}

In Fig.~\ref{fig:massvssfr}, we investigate the relation between the
stellar mass \citep[derived from rest-frame R-band magnitudes,
  see][for more details]{koy13a} and the star formation rate (derived
from the H$\alpha$ line) for the complete sample of HAEs discovered in
the field of MRC1138$-$262. In addition, we show the location of HAEs
counterparts of LABOCA sources that are protocluster members. There seems to be a weak trend that the HAEs associated with LABOCA sources are more massive and have higher
SFRs than the overall population of HAEs in the field
of the radio galaxy. In addition, we derive SFRs based on the far-IR measurements for LABOCA sources selected as HAEs and reveal that these sources (far-IR bright HAEs) are off the star-formation main sequence for $z\sim2$ galaxies \citep{dad07,san09}.

\subsection{Large scale structure traced by SMGs at $z=2.2$}
\label{sec:lsssmg}
We compare the number counts of the MRC1138$-$262 field with APEX
LABOCA blank field observations of the LESS survey \citep{wei09}, extracting sources with
a significance level above 3.7$\sigma$ (the extraction limit of the LESS survey) in our data. We detect six sources with
at least 3.7$\sigma$ and fluxes greater than 7.0~mJy in our map (DKB01, DKB02, DKB03, DKB05, DKB09, DKB11). The
size of the map with a maximum noise of 1.9~mJy is 56~arcmin$^2$,
yielding an estimated surface density of
0.107~arcmin$^{-2}$. According to the number counts derived by
\citet{wei09}, we expect a surface density down to the flux level of
7~mJy of 0.028~arcmin$^{-2}$. However, we should take into account  the fact that the LESS field could be underdense compared to previous submm surveys \citep[e.g.,][]{wei09,war11} down to $S_{850~\mu m}\gsim3$~mJy by a factor of two. Thus the source density in the
MRC1138$-$262 field is approximately two (to four) times higher than expected.

We compare the differential source counts for our sources with $S_{850\mu m}\gsim7$~mJy
to those in the ECDFS \citep{wei09} and corrected (multiplied) by the 'underdense factor' of two \citep[see][for this approach]{swi14}. We find that we can fit the
differential source counts in our field very well with the curves
fitted to the ECDFS counts, as provided by \citet{wei09}, normalised by
a factor 3.8. This implies that over the (small) range of source
fluxes probed by our map, we consistently find a 3.8 times higher density
as compared to a blank field. 

To understand how unusual the collection of SMGs found in the field of
MRC1138$-$262 is, we compared it with the spatial distribution of
sources in the ECDFS \citep{wei09}. We count the number of
$S_{850}>7$~mJy sources within 10,000 randomly placed circles each
having a radius of 4.22$^{\prime}$ (i.e., an area of 56~arcmin$^{2}$)
in the ECDFS. The most common number of sources found is one (40\%),
followed by zero (33\%). The highest number of sources found is six, and
95\% of the regions contain four sources or fewer. This means that the
surface density we measure in the field of MRC1138$-$262 is unusually
high and we cannot find a similarly dense field in the entire ECDFS
(almost 900 square arcmin).

To summarize, a comparison of the number counts suggests that we have
detected a significant excess of SMGs in the field of
MRC1138$-$262. However, is this overdensity connected to the
protocluster structure at $z\approx2.2$? In the following we discuss the evidence that the discovered overdensity is indeed associated with the protocluster at $z\approx2.2$. Including the radio galaxy,
five of the 16 SMGs (DKB01, DKB03, DKB07, DKB12, DKB16) are
spectroscopically confirmed members of the protocluster at
$z\approx2.2$.  A further two SMGs (DKB02, DKB15) have
photometric redshifts that suggest they could be protocluster
members. Our data excludes possible protocluster membership for three
sources (DKB06, DKB13, DKB14). For the remaining six sources (DKB04,
DKB05, DKB08, DKB09, DKB10, DKB11), we cannot make a judgement on
protocluster membership based on the data in hand. At least seven and up to 13 SMGs belong to the protocluster at $z\approx2.2$.

All six spectroscopically confirmed SMG members of the protocluster structure at $z\approx2.2$ --- five LABOCA sources plus the CO-bright HAE associated with SCUBA emission --- 
are located within a circle of 240$^{\prime\prime}$ diameter,
corresponding to 2.0~Mpc. In addition, both of the possible
members, DKB02 and DKB15 also lie within this area. Calculating the
surface density in this area as before (three sources fulfilling the
flux density limit of 7.0~mJy and detection level of 3.7$\sigma$ \citep[following][]{wei09}, two of them are spectroscopically confirmed protocluster
members), we derive a surface density of 0.239~arcmin$^{-2}$, a factor
of 4.3 higher than expected in a blank field at this wavelength.

Assuming a sphere of 2~Mpc, we calculate a SFRD of
$\sim$1500~M$_{\odot}$~yr$^{-1}$~Mpc$^{-3}$ which is four orders of
magnitude greater than the global SFRD at this redshift
\citep{hop06}. The SFRD of our protocluster is similar to results
obtained by \citet{cle14} for two clumps of HerMES sources at $z=2$.

The detection of an overdensity of SMGs at $z\approx2.2$ is consistent
with the overdensity of {\it Herschel} SPIRE 500~$\mu$m sources found
by \citet{rig14}. We note that none of our LABOCA sources are located
in the region where \citet{val13} reported an excess of
SPIRE~250~$\mu$m sources at 5$\sigma$ at a similar redshift 7$^{\prime}$ south of
the protocluster structure. Several groups have previously found
excesses of SMGs near HzRGs and QSOs
\citep[e.g.,][]{ivi00,ste03,deb04,gre07,pri08,ste10,car11}. In
comparison to our work, none of them have direct probes that a
significant fraction of their sources also lie at the redshift of the
targeted HzRG or QSO.

 \citet{bla04} report an association of five sources in the
 HDF-North. All five SMGs have spectroscopically measured redshifts of
 $z=1.99$ \citep[see also][]{cha09}. This is the largest blank field
 SMG association known so far. It is spatially distributed on a larger
 region on the sky than the MRC1138$-$262 group, spanning a region of
 7$^{\prime}$ ($\sim$3.5~Mpc) on a side. \citet{cha09} report an
 apparently less significant overdensity of UV-selected galaxies at
 the same redshift and region of the sky. Another association of three
 SMGs lies in the same field but at $z\approx4.0$
 \citep{dad09a,dad09b}. To summarize, the protocluster at
 $z\approx2.2$ is securely traced by galaxy populations probing
 different mass ranges, star formation and degree of obscuration
 including LAEs, HAEs, EROs and SMGs
 in the protocluster. \citet{rig14} observe several known protocluster structures with
SPIRE but do not recover SMG overdensities  for many of them. On average they detect more SPIRE sources than
compared to a blank field, and they 
 detect an overdensity of SPIRE 500~$\mu$m sources in the MRC1138$-$262 field. Focusing on HyLIRGs selected from {\it Herschel} wide field imaging, \citet{ivi13}
discovered a cluster of star-bursting proto-ellipticals at
$z=2.41$. \citet{sma14} related 31 FIR-/submm-selected sources to the $z=1.62$ cluster Cl0218.3$-$0510. Contrarily, \citet{bee08} report APEX LABOCA observations
of the J2142$-$4423 Ly$\alpha$ protocluster at $z=2.38$ and do not
find an excess of SMGs in this field. Similarly, {\it Herschel} SPIRE
observations by \citet{wyl13} do not confirm the previously reported SMG
overdensity in the field of 4C$+$41.17 \citep{ivi00}. Overall, there is significant evidence both from our work and from the literature that the detection of large scale structures in the early universe by far-infrared/submm observation are feasible but still not common.

\begin{figure*}[!]
 \centering
 \includegraphics[width=15cm,angle=0]{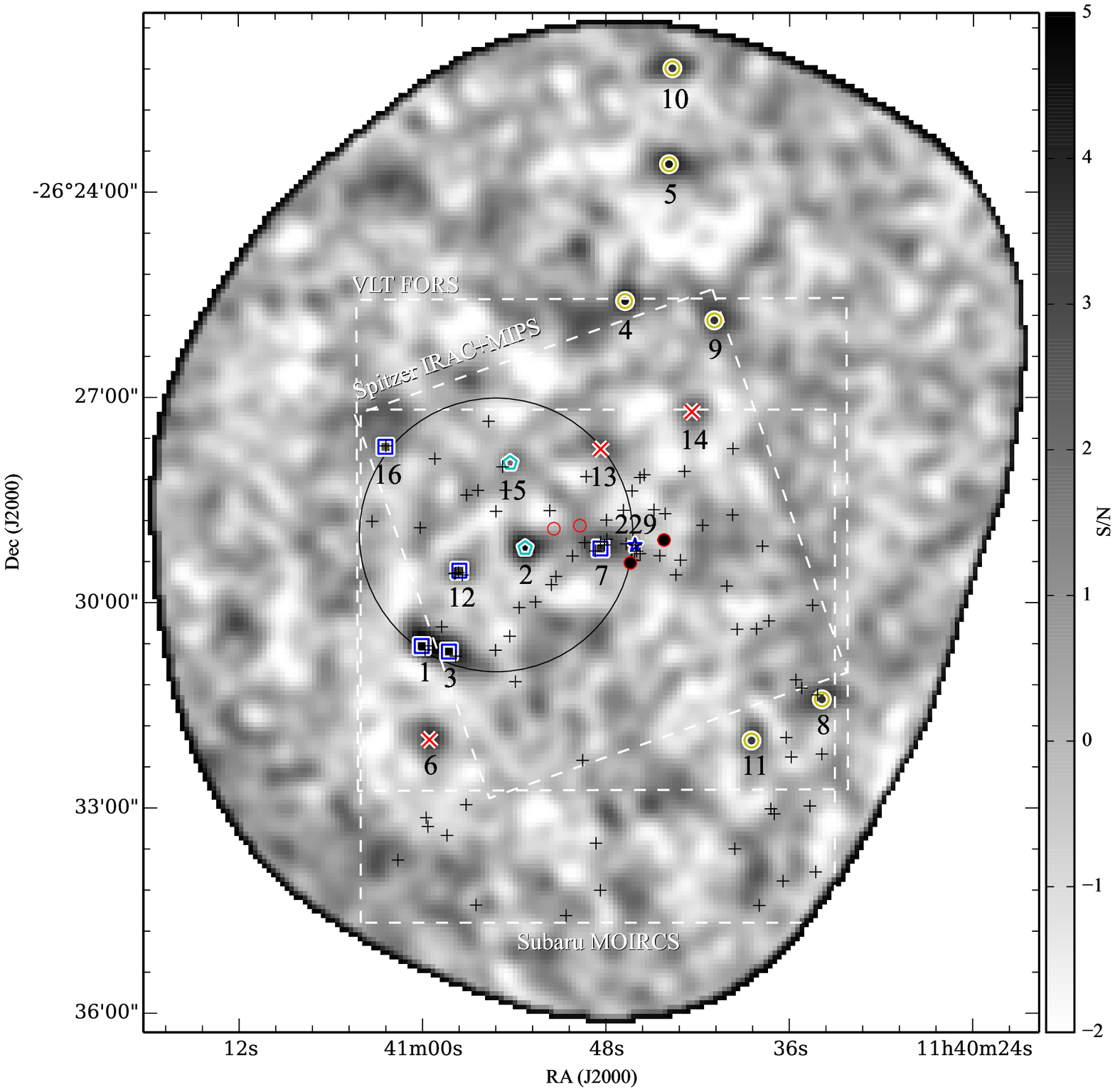}
 \caption{The location of 16 SMGs extracted from our LABOCA
   map of the field of MRC1138 on top of the LABOCA signal-to-noise
   map. Blue squares represent spectroscopically confirmed membership to the protocluster
   structure at $z\approx2.2$. The blue star is the SCUBA source at $z=2.149$, also detected in CO(1-0) by \citet{emo13}. Cyan pentagons show possible protocluster
   members. In the case of yellow circles, no reliable judgment on the
   cluster membership can be made. Red crosses are sources that can be
   securely excluded from the protocluster. The large circle has a
   diameter of $\sim$240$^{\prime\prime}$ (corresponding to a physical
   size 2~Mpc) and shows the region where all eight SMGs at $z=2.2$
   are located. The SMG overdensity is at least a factor four higher than compared
   to blank fields \citep{wei09} and not centered on the radio galaxy
   MRC1138 (DKB07). The spatial distribution of the SMG overdensity seems to be similar to the north-east and south-east filament-like structure traced by HAEs \citep[plus
     symbols][]{kur04b,koy13a} and in contrast to the location of passive
   quiescent galaxies clustered within 0.5~Mpc around the radio galaxy
   \citep[red circles, filled if spectroscopically
     confirmed, see][]{tan13}. In addition, we show the fields of view of
   our {\it Spitzer} IRAC/MIPS, VLT FORS and Subaru MOIRCS datasets. North
   is at the top and east is to the left. }
 \label{fig:member}
\end{figure*}

\citet{koy13a} find a clustering of HAEs around the radio galaxy
MRC1138$-$262 and report a large filament from north-east to
south-west ($\gsim10$~Mpc); a part of this filament was seen in the data of  \citet{kur04b}.  The SMGs belonging to the protocluster at $z\approx2.2$ are distributed within the north-east filament \citep{kur04b,koy13a} and the possible extension to the south-east \citep{koy13a} but not within the filament towards the south-west. However, due to the low number statistic, we cannot make a firm statement if the cosmic web could be traced by our SMGs. Our SMG overdensity is not
centered on the radio galaxy, which lies at the western edge of the SMG
concentration (see Fig.~\ref{fig:member}). A radial source density analysis strengthens this finding. This is in contrast to
the four passive quiescent galaxies which cluster within
0.5~Mpc of the radio galaxy \citep{tan13}. The following hypothesis could explain these findings: both populations are massive but
those in the centre have lost their gas; those that are still infalling have gas and are possibly being disturbed which makes them more
active \citep[see e.g.][for details on this kind of scenario at low redshift]{ver12}. The
HAE and SMG centres seem to be inconsistent, see
Fig.~\ref{fig:member}. This finding is similar to that of
\citet{koy13b} for a cluster at $z=0.4$ where a higher SFR is measured
by the IR in the cluster center whereas the SFR derived from H$\alpha$ is
similar to that of the field. Their conclusion is that the dust
extinction in galaxies in high density regions is higher than those in
the field, at the same redshift. 

\section{Conclusions}
We have mapped the field of MRC1138$-$262 ($\sim$140 arcmin$^{2}$)
with APEX LABOCA at 870~$\mu$m. This field has an exquisite
multiwavelength dataset, close in quality to the ECDFS, including
optical-NIR (VLT and Subaru), {\it Herschel} PACS+SPIRE, {\it Spitzer} IRAC, MIPS
24~$\mu$m, and deep HST and VLA 1.4~GHz imaging, as well as VLT FORS2, ISAAC and
SINFONI spectroscopy of protocluster members. 

\spb In total, we detected 16 SMGs --- 12 solid 3.5$\sigma$ and 4 cross-identified tentative detections --- with flux densities in the range 3--11~mJy. This is approximately a 
 factor up to four more than expected from blank field surveys such as LESS  \citep{wei09}, --- based on six sources with $S_{870~\mu m}>7$~mJy and $>3.7\sigma$ significance level. This excess is
consistent with the excess of SPIRE 500~$\mu$m sources found by \citet{rig14} at larger scales.

\spb Based on VLA 1.4~GHz, {\it Herschel}, {\it Spitzer} MIPS and Subaru
rest-frame H$\alpha$ imaging at $z\sim2.2$, we have identified the counterparts of
the LABOCA sources and derived reliable far-infrared photometric
redshifts. 55\% of the SMGs with $z\approx2.2$
H$\alpha$ imaging coverage are associated with HAEs. Near-infrared spectroscopic observations with VLT ISAAC and
SINFONI have confirmed redshift to be $z=2.16$ for four of these
SMG counterparts. Including the radio galaxy, five out of 16 SMGs are secure protocluster members at
$z\approx2.2$. Another two SMGs have photometric redshifts suggesting that they are possible protocluster members. Our data excludes  the protocluster membership
for three SMGs. For the remaining six
SMGs we do not have enough data
to make a robust judgement on their protocluster membership. 
 
\spb We associate the spectroscopically confirmed HAE229 \citep{kur04b} at $z=2.149$, recently detected in CO(1-0) \citet{emo13}, with a SCUBA source \citep{ste03}. This source is detected in {\it Herschel} bands and the far-IR photo-z is consistent with its spectroscopic redshift. Thus, we conclude that this CO-bright HAE is an SMG related to the protocluster at $z\approx2.2$, increasing the number of spetroscopically confirmed SMGs as protocluster members to six. 

\spb All six spectroscopically confirmed members of the protocluster structure at $z\approx2.2$ are
located within a circle of $\sim$240$^{\prime\prime}$ diameter,
corresponding to 2.0 Mpc at this redshift. Both of the possible members, DKB02 and DKB15, also lie within this area. The excess of SMGs in this region is at least four times higher than expected from 
blank fields. For comparison, the surface density of LABOCA sources is
significantly higher than the well known structure of six SMGs at
$z=1.99$ in GOODS-N distributed over 7$\times$7 Mpc$^{2}$
\citep{bla04,cha09}. The SMG overdensity is not
centered on the radio galaxy, which lies at the edge of the dusty
starburst concentration. The spatial distribution of the SMG overdensity seems to be similar to the north-east and south-east filament-like structure traced by HAEs \citep{kur04b,koy13a}. The SFR$_{FIR}$ of the LABOCA sources related to the protocluster  ranges between 200 to 1800 M$_{\odot}$~yr$^{-1}$ and sums up to a star formation rate
    density SFRD~$\sim$1500~M$_{\odot}$~yr$^{-1}$~Mpc$^{-3}$, four magnitudes
    higher than the global SFRD at this redshift in the field. 

Our results demonstrate that submillimeter observations can reveal clusters of massive, dusty starbursts. We show that at submm wavelengths systematic and detailed investigations of distant clusters are possible. However, we emphasize that only sensitive subarcsecond resolution
observations with ALMA will allow a complete characterization of the
16~SMGs discovered by LABOCA.

\begin{acknowledgements} 
Based on observations made with ESO Telescopes at Chajnantor and
Paranal under programme 084.A-1016(A), 083.F-0022, 088.A-0754(A) and
090.B-712(A). This work is based on observations with the APEX
telescope. APEX is a collaboration between the Max-Planck-Institut
f\"ur Radioastronomie, the European Southern Observatory, and the
Onsala Observatory. We are very grateful to Ian Smail who encouraged us to carry out this project and gave helpful advice during the project. We are much obliged to instructive help by Bjorn Emonts regarding the CO(1-0) observations of HAE229. We
would like to thank the APEX staff for their support during the
observations and Chris Carilli for his help during the VLA data
reduction. The National Radio Astronomy Observatory is a facility of the National Science Foundation operated under cooperative agreement by Associated Universities, Inc. We also acknowledge the contribution by the anonymous referee in clarifying a number of important points and thus improving this manuscript. We are grateful to Elaine Grubmann for proofreading. This publication is supported by the Austrian Science Fund (FWF). NS is supported by an ARC Future Fellowship. {\it Herschel}
is an ESA space observatory with science instruments provided by
European-led Principal Investigator consortia and with important
participation from NASA. PACS has been developed by a consortium of
institutes led by MPE (Germany) and including UVIE (Austria); KU
Leuven, CSL, IMEC (Belgium); CEA, LAM (France); MPIA (Germany);
INAF-IFSI/OAA/OAP/OAT, LENS, SISSA (Italy); IAC (Spain). This
development has been supported by the funding agencies BMVIT
(Austria), ESA-PRODEX (Belgium), CEA/CNES (France), DLR (Germany),
ASI/INAF (Italy), and CICYT/MCYT (Spain). SPIRE has been developed by
a consortium of institutes led by Cardiff University (UK) and
including Univ. Lethbridge (Canada); NAOC (China); CEA, LAM (France);
IFSI, Univ. Padua (Italy); IAC (Spain); Stockholm Observatory
(Sweden); Imperial College London, RAL, UCL-MSSL, UKATC, Univ. Sussex
(UK); and Caltech, JPL, NHSC, Univ. Colorado (USA). This development
has been supported by national funding agencies: CSA (Canada); NAOC
(China); CEA, CNES, CNRS (France); ASI (Italy); MCINN (Spain); SNSB
(Sweden); STFC (UK); and NASA (USA).
\end{acknowledgements}

\end{document}